%% file: main.tex
\documentclass{article}
\PassOptionsToPackage{dvipsnames}{xcolor}
\IfFileExists{iclr2027_conference.sty}%
  {\usepackage{iclr2027_conference,times}}%
  {\usepackage{iclr_fallback,times}}

\usepackage[utf8]{inputenc}
\usepackage[T1]{fontenc}
\usepackage{microtype}

\usepackage{amsmath,amssymb,amsfonts}
\usepackage{bm}
\usepackage{nicefrac}

\usepackage{xcolor}
\usepackage{colortbl}

\usepackage{booktabs}
\usepackage{graphicx}
\usepackage{multirow}
\usepackage{adjustbox}
\usepackage{caption}
\usepackage{subcaption}
\usepackage{tabularx}
\usepackage{makecell}
\usepackage{array}

\usepackage{enumitem}
\usepackage{algorithm,algorithmic}
\makeatletter
\renewcommand\fs@ruled{%
  \def\@fs@cfont{\bfseries}\let\@fs@capt\floatc@ruled
  \def\@fs@pre{{\color{black}\hrule height.8pt depth0pt}\kern2pt}%
  \def\@fs@post{\kern2pt{\color{black}\hrule}\relax}%
  \def\@fs@mid{\kern2pt{\color{black}\hrule}\kern2pt}%
  \let\@fs@iftopcapt\iftrue}
\makeatother
\floatstyle{ruled}\restylefloat{algorithm}

\usepackage{hyperref}
\usepackage{url}
\hypersetup{colorlinks=true}
\hypersetup{linktoc=all}
\hypersetup{citecolor=customblue}
\hypersetup{linkcolor=crimson}
\hypersetup{urlcolor=MidnightBlue}
\usepackage[all]{hypcap}
\usepackage[nameinlink]{cleveref}

\definecolor{crimson}{rgb}{0.86, 0.08, 0.24}
\definecolor{customblue}{rgb}{0.2, 0.35, 0.8}
\definecolor{keyrow}{gray}{0.92}

\newcommand{\hi}{\rowcolor{keyrow}}

\newcommand{\kvcomm}{\textsc{KVComm}}
\newcommand{\ogate}{\textsc{OpenGate}}  
\newcommand{\repair}{\textsc{BCR}}  
\newcommand{\bcrchunk}{\textbf{\textsc{BCR}-chunk}}     
\newcommand{\bcrgrow}{\textbf{\textsc{BCR}-grow}}       
\newcommand{\bcrsingle}{\textbf{\textsc{BCR}-single}}   
\newcommand{\dense}{\textsc{Dense}}  
\newcommand{\gate}{\gamma}

\renewcommand{\algorithmiccomment}[1]{\bgroup\hfill$\triangleright$~#1\egroup}

\title{Budgeted Cache Repair\\
for Cross-Context \mbox{KV-Cache} Reuse}

\iclrfinalcopy
\author{%
\mdseries 
\begin{tabular}[t]{@{}l@{\hspace{5em}}l@{}}
Haeyong Kang\thanks{Corresponding author: hykang@duksung.ac.kr.} & Chang D. Yoo \\
Duksung Women's University & KAIST \\ & \\
\end{tabular}}

\begin{document}

\maketitle
\suppressfloats[t]
\fancyhead[L]{Preprint. Under review.}
\thispagestyle{fancy}

\input{sections/abstract}
\input{sections/introduction}
\input{sections/related_work}

\input{sections/bcr}

\input{sections/setup}

\input{sections/audit}

\input{sections/allocation}

\input{sections/grains}

\input{sections/conclusion}
\input{sections/statements}

\bibliography{references}
\IfFileExists{iclr2027_conference.bst}%
  {\bibliographystyle{iclr2027_conference}}%
  {\bibliographystyle{plainnat}}

\appendix
\input{sections/method}
\input{sections/selective}
\input{sections/experiments}
\input{sections/cacheblend}
\input{sections/abovedense}
\input{sections/costsummary}
\input{sections/appendix}
\input{sections/limitations}

\end{document}

%% file: sections/abstract.tex
\begin{abstract}
Cross-context KV-cache reuse predicts a shared segment's keys and values under a new prefix instead of recomputing them, and has been reported to do so without quality loss. We find otherwise, and identify two problems. (1) \textbf{A hidden cost}: on MMLU and GSM8K, reuse costs substantial accuracy. (2) \textbf{A decision at the wrong unit}: no rule for deciding \emph{whether} to reuse a cache removes that cost. What does help is choosing \emph{which parts} of the cache to recompute, and the value of choosing well falls as the unit of choice grows: informed selection removes $49.5\%$ of the cache error beyond chance at single rows (one token's keys and values), $10.6\%$ at $64$-token chunks, and nothing at the level of whole calls. \textbf{B}udgeted \textbf{C}ache \textbf{R}epair (\textbf{\repair{}}) acts at the unit where selection still pays. It drafts two tokens from the assembled cache, ranks cache rows by the attention those tokens pay them, and recomputes a fixed budget of rows exactly, in one of three layouts. The cost is paid rather than predicted away, and \emph{the draft that fails as a gate succeeds as a selector}. \repair{} restores GSM8K to dense-prefill accuracy while still serving most calls from cache, and its best layout outperforms every reuse baseline's mean in the reference grid. The draft also beats a coin-flip selector at the same budget---a control prior evaluations lack.
\end{abstract}

%% file: sections/introduction.tex
\section{Introduction}
\label{sec:intro}

A multi-agent LLM system processes the same context again and again. When three agents and an aggregator each receive the same question and each other's outputs, the question is prefilled four times and the first answer three times. This redundancy motivates cross-context KV-cache reuse \citep{kwon2023vllm,gim2024promptcache,yao2025cacheblend}: rather than recomputing a segment's keys and values under every new prefix, a reuse method estimates whether the representations computed earlier are still valid. \kvcomm{} \citep{ye2025kvcomm} does this with an online pool of \emph{anchors}---segments stored together with the deviation a prefix change induced in them---and reports over $70\%$ cache reuse and up to $7.8\times$ faster time-to-first-token with no measurable quality loss.

\input{sections/fig_concept}

A reused cache is not always safe to serve, so every reuse system needs a rule for deciding when its prediction can be trusted. CacheBlend \citep{yao2025cacheblend}, \Cref{fig:concept}(a), sidesteps the question by recomputing selected tokens instead of trusting the seams. \kvcomm{} instead applies a \emph{gate}: once per reused segment, before generation begins, it thresholds the entropy of the segment's similarity to the stored anchors (\S\ref{sec:bcr}). The gate's defining limitation, \Cref{fig:concept}(b), is that \emph{it decides before any information about the answer exists}: it can inspect only the prompt and the predicted cache, so it uses the cache's geometry as a surrogate for whether reuse will change the answer.

We find that this prompt-side surrogate is poorly aligned with the errors reuse actually causes. Reuse costs substantial accuracy on two of three benchmarks, and the cost survives a check for anchor-pool warm-up. Tightening the gate does not remove it, and neither does choosing which calls to reuse: across five call-level allocation policies, including a random baseline, accuracy tracks the amount of reuse rather than the policy. Nor does deferring the decision until generation has begun, the fix that cache eviction uses \citep{li2024snapkv,zhang2023h2o,ge2025lookahead}: our generation-time gate \ogate{}, which decides from two draft tokens, loses $6.7$ points to \dense{} on $300$ matched GSM8K problems (exact McNemar $p=0.009$ on the problems where the two disagree; \S\ref{sec:setup} gives the test form). Together these results say the problem is not \emph{when} or \emph{whether} to reuse but \emph{what unit} the decision controls: a whole-call gate can only trust or reject an entire cache, even when the error sits in a small subset of its rows.

This motivates a different approach: \emph{repair part of the cache instead of trusting all of it}. \textbf{B}udgeted \textbf{C}ache \textbf{R}epair (\textbf{\repair{}}) uses generation itself to find the parts of a reused cache most worth repairing. It drafts two tokens from the assembled cache, ranks cache rows by how much attention those tokens pay them, and recomputes exactly a fixed budget of the highest-ranked rows. We study three repair layouts, illustrated in \Cref{fig:concept}(c): fixed chunks, grown spans, and individual rows (\S\ref{sec:bcr}). \repair{} pays the hidden cost directly, and pays it at a unit where informed selection can still reduce it.

Our contributions are as follows.
\begin{itemize}[leftmargin=*]

\item \textbf{Budgeted Cache Repair.}
We introduce \repair{}, a generation-guided method that uses a two-token draft to select, and then exactly recompute, a fixed budget of KV rows, in three repair layouts: fixed chunks, grown spans, and individual rows.

\item \textbf{Accuracy recovery under reuse.} \repair{} recovers most of the accuracy reuse costs, restoring GSM8K to dense-prefill accuracy while keeping most of the reuse, and on each benchmark of the reference grid its best layout leads every reuse baseline's mean, \kvcomm{} and CacheBlend alike. Recovery is task-dependent: gains on MMLU are limited, and HumanEval has little degradation to recover.

\item \textbf{What determines the error of reuse.}
The value of informed selection falls as the unit of decision grows, from single rows to chunks to whole calls, and the cost itself is workload- or backbone-specific.

\end{itemize}

%% file: sections/fig_concept.tex
\begin{figure}[tbp]
\centering
\begin{subfigure}[b]{0.33\textwidth}
  \centering
  \includegraphics[width=\textwidth]{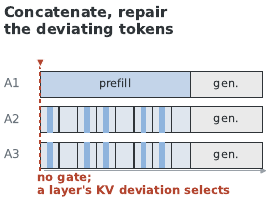}
\end{subfigure}\hfill
\begin{subfigure}[b]{0.33\textwidth}
  \centering
  \includegraphics[width=\textwidth]{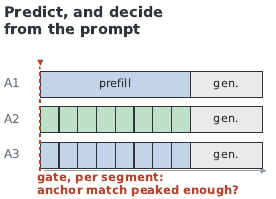}
\end{subfigure}\hfill
\begin{subfigure}[b]{0.33\textwidth}
  \centering
  \includegraphics[width=\textwidth]{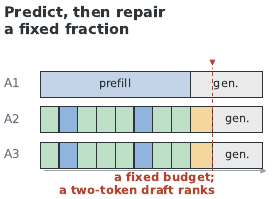}
\end{subfigure}

\vspace{0pt}
\includegraphics[width=0.95\textwidth]{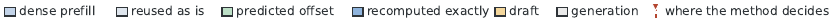}
\vspace{0pt}

\captionsetup[subfigure]{font=small,skip=2pt}
\begin{subfigure}[t]{0.305\textwidth}
  \caption{\textbf{CacheBlend.} No offset; a layer's deviation
  selects.}
  \label{fig:concept-cb}
\end{subfigure}\hfill
\begin{subfigure}[t]{0.305\textwidth}
  \caption{\textbf{\kvcomm{}.} Offset from anchors; a gate
  refuses A3.}
  \label{fig:concept-b}
\end{subfigure}\hfill
\begin{subfigure}[t]{0.305\textwidth}
  \caption{\textbf{\repair{} (ours).} Offset kept; a draft
  selects a budget.}
  \label{fig:concept-c}
\end{subfigure}

\caption{\textbf{Three ways to serve a shared segment.} Rows show three agents sharing one segment in execution order: A1 is the \emph{first caller} and computes it exactly; A2 and A3 are \emph{later callers} and can reuse it from cache. The horizontal axis denotes token position.}
\label{fig:concept}
\end{figure}

%% file: sections/related_work.tex
\section{Related Work}
\label{sec:related}

\paragraph{Reuse across contexts.}
Prefix caching \citep{kwon2023vllm} reuses a cache only under an exact prefix match. Prompt Cache \citep{gim2024promptcache} extends reuse to position-agnostic segments, and CacheBlend \citep{yao2025cacheblend}, \Cref{fig:concept}(a), repairs independently cached chunks by recomputing selected tokens. \kvcomm{} \citep{ye2025kvcomm}, \Cref{fig:concept}(b), takes a different route: it predicts the deviation a prefix change causes, using an online anchor pool, and decides per segment whether to reuse the cache. We study this formulation and ask whether such prediction preserves accuracy, including after the anchor pool has warmed up. Multi-agent pipelines \citep{du2023debate,zhuge2024gptswarm} make such reuse frequent; AgentPrune \citep{zhang2024agentprune} attacks the same cost from the communication side. Cache-compression methods face a related choice of what to keep, using positional rules \citep{xiao2024streamingllm}, reconstruction \citep{zweiger2026fast,kim2025kvzip}, or attention-based scores \citep{li2024snapkv,zhang2023h2o,cai2024pyramidkv,feng2024adakv}; recent work shows that such input-side criteria can be misaligned with task performance \citep{ge2025lookahead}. \repair{} differs from all of these in what it decides: it fixes a repair budget and chooses \emph{what to repair}, rather than whether to trust the cache as a whole. Like CacheBlend it pays for selective recomputation, but it applies the idea to cross-context reuse and studies how the unit of repair changes its value; \Cref{fig:concept}(c) draws the contrast.

\paragraph{Selecting what to recompute.}
Recent work improves selective recomputation by ranking tokens with attention from the user query \citep{teng2026infoflow,wang2026prophetkv} or from a small model that approximates the target \citep{yang2025cacheclip}, and reports substantial gains at matched budgets. Draft passes that generate tokens before committing \citep{leviathan2023speculative,ge2025lookahead} likewise motivate generation-time signals for selection. These methods ask \emph{which tokens to select} within a fixed formulation. We isolate two factors instead: what informed selection is worth relative to an uninformed baseline, and how that worth depends on the \emph{unit} of selection. We therefore compare informed and random selection at the same repair budget, and vary the unit from whole calls to chunks to individual cache rows. That controlled view is the focus of \repair{}.

%% file: sections/bcr.tex
\section{Budgeted Cache Repair}
\label{sec:bcr}

\paragraph{The reuse problem.}
A request runs on a directed acyclic graph of $N$ agents. Agent $i$'s prompt is assembled from a private system prompt $S_i$, the shared task $T$, and the outputs of its predecessors, so in the fully connected configuration studied here a segment such as $T$ is prefilled $N{+}1$ times. Let $C(x \mid P)$ denote the KV cache of a token span $x$ computed under prefix $P$, one \emph{row} per token. Every entry of that cache depends on $P$ through attention, so exact reuse requires an identical prefix---the condition prefix caching assumes \citep{kwon2023vllm}. \kvcomm{} \citep{ye2025kvcomm} relaxes it by \emph{predicting} the dependence. For a segment $x$ already seen under some prefix $P$, it stores an \emph{anchor}: the deviation
\begin{equation}
\Delta \;=\; C(x \mid P) \;-\; C(x \mid \emptyset),
\label{eq:delta}
\end{equation}
between the segment's cache under $P$ and its cache under no prefix at all. At serve time, a new segment's deviation is approximated by a similarity-weighted combination of stored $\Delta$'s, added elementwise to the cache.

\paragraph{The gate.}
An agent's prompt contains several such segments---the task, each predecessor's output---and each is decided separately: trust the approximation (\emph{reuse}) or pay a full prefill for that segment (\emph{dense}). \kvcomm{} decides \emph{before generating any token}, from an input-side statistic. Let $d_j$ be the $L_2$ distance in value space between the candidate segment and anchor $j$, collect the distances over the $n$ available anchors as $d = (d_1,\dots,d_n)$, and let $s = \mathrm{softmax}(-d)$. The rule thresholds the entropy of $s$:
\begin{equation}
\text{reuse}
\quad\Longleftrightarrow\quad
H(s) \;\le\; \gate \log_2 n,
\qquad
H(s) = -\textstyle\sum_j s_j \log_2 s_j ,
\label{eq:gate}
\end{equation}
with a hard fallback to dense whenever $n \le 1$. A peaked $s$ means the candidate closely matches one stored anchor, whose deviation should transfer. Since $\log_2 n$ is the largest entropy $s$ can have, the threshold $\gate$ (default $0.3$) is the fraction of maximum ambiguity the rule tolerates, and $\gate\to 1$ recovers unconditional reuse. \Cref{eq:gate} is the object this paper examines.

\paragraph{The repair.}
Predicting a cache instead of computing it costs accuracy, and no setting of the gate avoids that cost; \S\ref{sec:audit} measures both. Reusing fewer caches, and choosing which, does not avoid it either: \S\ref{sec:allocation} rules that out and prices what choosing well is worth at each unit. What remains is to repair part of the reused cache rather than decide whether to trust all of it. As \Cref{fig:concept}(c) illustrates, \repair{} recomputes a fixed \emph{budget} $\rho$ of the cache's rows exactly and reuses the rest. Whenever \kvcomm{} serves a call from an assembled cache, \repair{} generates two tokens, rolls the draft back, and scores every cached row by the attention the draft paid it. Let $Q^{\ell}$ be the two query rows kept at layer $\ell$, indexed by step $t$ and head $h$, and $\widetilde{K}^{\ell}$ the predicted keys. Row $j$ scores
$a_j = \sum_{\ell>1} \operatorname{mean}_{t,h} \big[\operatorname{softmax}(Q^{\ell}_{t,h}\widetilde{K}^{\ell\top}_{h}/\sqrt{d_{\mathrm{head}}})\big]_j$,
and a group of rows scores $\bar a_i = \operatorname{mean}_{j\in i} a_j$. The highest-scoring rows or groups, up to a budget of $\rho$ of the cached rows, are recomputed exactly before generation begins. The keys scored are the predicted ones: the signal that decides what to fix is read from the tensors under suspicion. The draft is the same deferral \ogate{} applies to \emph{whether} to reuse, applied instead to \emph{which rows} to repair; App.~\ref{app:selective} gives the procedure as \Cref{alg:draft}.

\input{sections/table_budget}

\paragraph{The layouts.}
Spending a budget at all recovers $7.9$ of the $13.7$ accuracy points reuse costs on GSM8K, and the rule that places the budget recovers $5.4$ more (\S\ref{sec:grains}). How large the budget should be, \Cref{tab:budget}(a)'s coin-drawn sweep cannot resolve: at $\rho{=}0.20$ its seeds span $11.3$ accuracy points, wider than the $82.7-77.3$ gap they bracket. What it leaves open is the layout the budget is spent in, and we test three, coarsest first:

\begin{itemize}[leftmargin=*,itemsep=1pt,topsep=3pt,parsep=0pt]
\item \bcrchunk{} divides the cache into fixed $64$-token chunks and recomputes the highest-scoring $\rho$ of them. Spans have a fixed length and sit on a grid.
\item \bcrgrow{} drops the fixed unit: it seeds on the best-scoring row and extends the span outward while a neighbour scores at least $\tau$ of the seed. Lengths follow the score surface, and positions are not aligned to anything.
\item \bcrsingle{} drops grouping altogether and spends the budget one row at a time, the finest unit the budget admits.
\end{itemize}

\paragraph{One budget, three units.}
Unqualified, \repair{} means all three. \Cref{tab:budget}(b) puts their row counts within $12\%$ of the smallest ($154$ against $138$ on HumanEval, the widest of the three; $9.4\%$ on MMLU and $4.7\%$ on GSM8K), so the three share one budget and differ in the unit they spend it in; comparing them isolates the unit, not the spend. \Cref{tab:budget}(c) prices the budget in time: $190.1$\,ms to the first token, against reuse's $72.9$ and \dense{}'s $542.7$.

%% file: sections/table_budget.tex
\begin{table}[!t]
\centering
\caption{\textbf{The budget: how large, in what layout, at what price.} (a) two agents, $150$ matched GSM8K problems, $64$-token chunks drawn by coin at four seeds so $\rho$
is the only variable; \emph{mean} and \emph{spread} are the accuracy over those seeds and its range, $\mathrm{max}-\mathrm{min}$. \kvcomm{} is $\rho{=}0$, \dense{} $\rho{=}1$. (b) rows recomputed per agent call at four agents. (c) time to the first token on a $4096$-symbol synthetic segment at $\rho{=}0.20$, $\times$ the speedup against \dense{}. Full tables in \Cref{tab:rho,tab:repaircost,tab:breakeven}.}
\label{tab:budget}
\renewcommand{\arraystretch}{0.9}

\begin{minipage}[t]{0.26\textwidth}
\centering\footnotesize\setlength{\tabcolsep}{1.5pt}
\begin{tabular*}{\linewidth}[t]{@{\extracolsep{\fill}}lcc}
\toprule
$\rho$ & mean & spread \\
\midrule
$0.05$ & 76.5 & $3.3$ \\
$0.20$     & 77.5 & $11.3$ \\
$0.40$ & 78.8 & $2.7$ \\
\midrule
\kvcomm{} & 77.3 & -- \\
\dense{}  & 82.7 & -- \\
\bottomrule
\end{tabular*}

\vspace{3pt}
{\small (a) The budget swept, four coin seeds}
\end{minipage}\hfill
\begin{minipage}[t]{0.42\textwidth}
\centering\footnotesize\setlength{\tabcolsep}{2pt}
\begin{tabular*}{\linewidth}[t]{@{\extracolsep{\fill}}lccc}
\toprule
Method & MMLU & GSM8K & HEval \\
\midrule
\hi \bcrchunk{}  & 159 & 211 & 138 \\
\bcrgrow{}       & 174 & 221 & 152 \\
\bcrsingle{}     & 167 & 216 & 154 \\
\midrule
CacheBlend, reuse & 417 & 474 & 97 \\
CacheBlend, cost  & 158 & 206 & 132 \\
\bottomrule
\end{tabular*}

\vspace{3pt}
{\small (b) Rows recomputed per call, four agents}
\end{minipage}\hfill
\begin{minipage}[t]{0.28\textwidth}
\centering\footnotesize\setlength{\tabcolsep}{2pt}
\begin{tabular*}{\linewidth}[t]{@{\extracolsep{\fill}}lcc}
\toprule
Method & ms & $\times$ \\
\midrule
\dense{}        & 542.7 & -- \\
\kvcomm{}       & 72.9  & 7.63 \\
\hi \bcrchunk{} & 190.1 & 2.85 \\
\bottomrule
\end{tabular*}

\vspace{3pt}
{\small (c) Time to first token, $4096$ symbols}
\end{minipage}
\end{table}

%% file: sections/setup.tex
\section{Experiments}
\label{sec:experiments}
\label{sec:setup}

\paragraph{The measurement setting.}
GSM8K \citep{cobbe2021gsm8k} and MMLU \citep{hendrycks2021mmlu} run on Llama-3.1-8B-Instruct \citep{dubey2024llama} and are scored by exact match and letter matching respectively; HumanEval \citep{chen2021humaneval} runs on Qwen2.5-Coder-7B-Instruct \citep{hui2024qwen25} and is scored by unit tests. Two configurations recur, and every result is labelled with the one it uses. The \emph{reference} configuration follows the released implementation: roles cycle rather than stay pinned, and the shipped question budgets---$150$, $153$ and $160$ items---are run at $2$ to $5$ agents in \Cref{tab:fair}. Its three-\texttt{MathSolver} GSM8K cell, with $N{+}1=4$ calls per request, is where \S\ref{sec:audit} begins: $300$ problems, the first $60$ of which also serve to sweep $\gate$. The \emph{pinned} configuration assigns five agents fixed roles, over $300$ matched GSM8K and MMLU problems in \Cref{tab:twobench} or the $150$ problems of \Cref{tab:evidence}(a); pinning concentrates divergence at the entry node, where \S\ref{sec:audit} measures it. Statistical significance in this paper comes from these single runs over distinct items, never from \Cref{tab:fair}. Decoding is greedy throughout, with $512$ maximum new tokens, on one NVIDIA A100 80GB; anchor-pool hyperparameters stay at the repository defaults. App.~\ref{app:t1setup} gives the full environment and scoring rules.

\paragraph{The statistical protocol.}
Within each comparison, every condition runs the \emph{same} problems with the same decoding parameters, so every comparison is paired. We report the \emph{discordant} split $b/c$ before any $p$-value: $b$ counts problems solved only by the first-named condition, $c$ those solved only by the other, unless a table's caption fixes the orientation for its rows. An exact McNemar test on these paired disagreements then asks whether they favour one condition more often than chance would. The split is more informative than marginal accuracy when two conditions differ on only a handful of problems. Results are reported per agent count rather than pooled across the grid: the four cells reuse the same problems, so pooling their pairs would treat repeated measurements as independent and inflate the effective sample size fourfold. Pooled splits therefore appear only descriptively, never as tests; App.~\ref{app:withdrawn} records the ones we withdrew.

%% file: sections/audit.tex
\subsection{What Reuse Costs}
\label{sec:audit}

We measure two things in order: whether predicting the cache instead of recomputing it costs accuracy, and whether that cost can be blamed on the \emph{rule} that decides when to trust a prediction---the gate of \Cref{eq:gate}. Both measurements compare the same problems solved two ways, under the matched-pair protocol above.

\input{sections/table_n300}

\paragraph{The cost of reuse.}
\Cref{tab:n300} puts the cost at $-4.3$ points on $300$ matched GSM8K problems in the reference configuration ($p = 0.035$ on $33$ discordant pairs); pinned, the cost is $13.7$ (\S\ref{sec:grains}). Unconditional reuse loses $7.7$ ($p = 0.001$), and \ogate{}, the answer-side gate of App.~\ref{sec:method}, loses $6.7$ ($p = 0.009$). No setting of the gate, and no replacement we built for it, avoids the cost of predicting the cache; recomputing part of the prediction does (\S\ref{sec:bcr}), while changing the decision about the whole cache does not. At $\gate = 0.3$, reuse loses $23$ problems that \dense{} solves but \emph{gains} $10$ that \dense{} misses, \Cref{tab:n300}'s $23/10$: $33$ problems change outcome for a net change of only $13$, or $4.3$ points. The approximated caches produce different answers, not occasional slips.

\input{sections/fig_audit}

\paragraph{The warm-up effect.}
On the first $60$ problems---a plausible pilot---reuse is \emph{ahead} by $1.7$ points, the opposite of the cost just measured. \Cref{fig:audit}(a) resolves this block by block, and the pattern follows the method's own dynamics rather than noise. The first block has the run's lowest reuse rate, $0.537$, against $0.625$ to $0.679$ in later blocks, because a dense pass is what \emph{materializes} an anchor: the realized deviation of \Cref{eq:delta} needs the exact cache for its subtraction, so early requests build the pool rather than draw on it. Every later block reuses more, and every later block is negative. A short evaluation therefore samples mostly the regime before the cost appears.

\paragraph{The gate sweep.}
If the loss came from misplaced trust in predictions, tightening the rule should recover quality. \Cref{tab:n300} sweeps it, and \Cref{tab:gamma} sweeps it more finely on the first $60$ problems, with latencies. The setting $\gate = 1.0$ is not an extrapolation but the exact ``never refuse'' policy, since $H(s) \le \log_2 n$ makes the test unfireable there---which App.~\ref{app:gsweep} uses as a check that the setting took effect. Every setting stays below \dense{}, and the cost deepens as reuse rises: $-4.3$, $-6.7$ and $-7.7$ accuracy points, at $p = 0.035$, $0.009$ and $0.001$. The gate moves the operating point along the one reuse--accuracy line that \Cref{fig:audit}(b) draws; it does not lift the line. The loss lives in the approximation itself, not in the rule that decides when to apply it.

\input{sections/table_fair}

\paragraph{The agent-count grid.} \Cref{tab:fair} repeats the comparison at two to five agents, and \dense{} leads \kvcomm{} in all eight MMLU and GSM8K cells. \emph{No single cell is significant} at the repository's own sample sizes: \Cref{tab:evidence}(d) gives all twelve splits; the largest, GSM8K at two agents, is $17/9$, $p = 0.169$. Pooling the counts is not a valid test (\S\ref{sec:setup}), so the grid gives direction, not magnitude; significance rests on the single $n{=}300$ runs.

The same grid puts the repair at \dense{}'s ceiling, where the reuse baselines stop short. In \Cref{tab:fair}'s means---descriptive, since the four cells re-run the same problems---\bcrchunk{} takes GSM8K at $84.7$, and \bcrgrow{} takes MMLU at $75.3$ and HumanEval at $88.4$: level with \dense{}'s $84.5$, $75.0$ and $86.1$, and clear of \kvcomm{}'s $80.8$, $72.5$ and $85.8$ and of CacheBlend's best, $83.7$, $71.7$ and $86.4$. The three layouts spend one budget, and which of them leads depends on the benchmark (\S\ref{sec:grains}); some layout beats \kvcomm{} in all twelve cells and each CacheBlend pairing in eleven.

The comparison matches conditions on budget, and \Cref{tab:fair}'s GSM8K cell re-runs to the printed scores exactly. MMLU does not. Its analysis agent splices a live, uncached \texttt{en.wikipedia.org} fetch into the prompt, so every condition---\dense{} included---depends on what Wikipedia said that day. Disabling retrieval makes the pipeline exactly reproducible, and that offline regime, documented in App.~\ref{app:remeasure}, is what \Cref{tab:fair}'s MMLU column now reports. Measured offline, the cost is $2.0$ to $2.7$ points---three or four problems of $153$ in \Cref{tab:evidence}(d)'s counts, none significant. The direction is unchanged, and the caveat applies to any MMLU number measured with roles cycling, the reference implementation's included.

\paragraph{The third benchmark.}
HumanEval behaves differently: reuse costs nothing. Its \kvcomm{} row runs at $\gate{=}1.0$, the setting whose reuse rate matches the published one, and across four cells it is level with \dense{} at reuse above $84\%$, $28/26$ in \Cref{tab:evidence}(d)'s pooled column. The tokenization term that \S\ref{sec:grains} prices at $6.7$ points on MMLU from \Cref{tab:twobench}(a) is measured there, not here, so we do not decompose this null. The cost is thus workload- or backbone-specific: present on two benchmarks, absent on the third. Where it is present, no rule avoids it---the premise from which \S\ref{sec:bcr}'s budget starts.

%% file: sections/table_n300.tex
\begin{table}[!t]
\centering
\caption{\textbf{What the gate can and cannot do.} The first $300$ GSM8K
problems in the reference configuration. $b$ counts problems \dense{} solves
and the row misses, $c$ the reverse, \emph{disc.}\ their sum; $^{\ast}$ marks
$p<0.05$. \emph{cost} is
points against \dense{}, taken from the discordant counts rather than the
rounded accuracies. \emph{rate} is the fraction of the $1200$ agent calls served
without a dense prefill.}
\label{tab:n300}
\small\setlength{\tabcolsep}{4pt}
\begin{tabular}{llccrccrr}
\toprule
& & \multicolumn{6}{c}{Accuracy vs.\ \dense{}} & Reuse \\
\cmidrule(lr){3-8}\cmidrule(lr){9-9}
Method & reuse when & acc.\ (\%) & 95\% CI & cost & $b/c$ & disc. & $p$ & rate \\
\midrule
\dense{}     & never & 85.7 & {\scriptsize [81.2, 89.2]} & -- & -- & -- & -- & 0.000 \\
\kvcomm{}    & $H(s) \le \gate{=}0.3\log_2 n$ & 81.3 & {\scriptsize [76.5, 85.3]} & $-4.3$ & 23/10 & 33 & 0.035$^{\ast}$ & 0.638 \\
\ogate{} & conf.\ $\ge \theta{=}0.5$ & 79.0 & {\scriptsize [74.0, 83.2]} & $-6.7$ & 37/17 & 54 & 0.009$^{\ast}$ & 0.902 \\
\kvcomm{} & $H(s) \le \gate{=}1.0\log_2 n$ & 78.0 & {\scriptsize [73.0, 82.3]} & $-7.7$ & 35/12 & 47 & 0.001$^{\ast}$ & 0.959 \\
\bottomrule
\end{tabular}
\end{table}

%% file: sections/fig_audit.tex
\begin{figure}[!tbp]
\centering
\begin{subfigure}[t]{0.48\textwidth}
  \includegraphics[width=\textwidth]{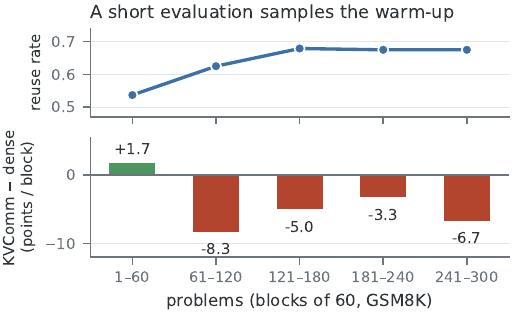}
  \caption{\textbf{The anchor-pool warm-up.} The pool fills as the run goes on,
  so the first block reuses least and is the only one where reuse does not
  lose.}
  \label{fig:res-warmup}
\end{subfigure}\hfill
\begin{subfigure}[t]{0.48\textwidth}
  \includegraphics[width=\textwidth]{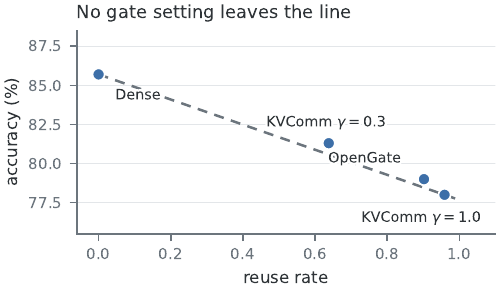}
  \caption{\textbf{The gate sweep.} Every setting, and the answer-side gate,
  lands on the line \dense{} and unconditional reuse define.}
  \label{fig:res-gate}
\end{subfigure}
\caption{\textbf{What reuse costs, and what the gate can do about it.} GSM8K,
$n{=}300$ (\Cref{tab:n300,tab:warmblocks}).}
\label{fig:audit}
\end{figure}

%% file: sections/table_fair.tex
\providecommand{\dummyhl}{}\definecolor{tabhl}{RGB}{232,238,246}
\begin{table}[!tbp]
\centering
\caption{\textbf{What reuse costs across agent counts.} Reference configuration,
$\gate{=}0.3$ ($1.0$ on HumanEval and on GSM8K's CacheBlend rows); $153$, $150$ and $160$ items, matched per problem.
\textbf{Bold} is the best score in a column excluding \dense{}, shading the best \repair{} mean; \emph{avg} carries no test. \dense{} reuses nothing and the cost-matched
pairing matches rows, so neither has a reuse row. Splits in
\Cref{tab:fairfull,tab:graincompare}.}
\label{tab:fair}
\scriptsize
\setlength{\tabcolsep}{3.2pt}\renewcommand{\arraystretch}{0.9}
\begin{tabular}{l cccc|c cccc|c cccc|c}
\toprule
& \multicolumn{5}{c}{MMLU} & \multicolumn{5}{c}{GSM8K}
& \multicolumn{5}{c}{HumanEval} \\
\cmidrule(lr){2-6}\cmidrule(lr){7-11}\cmidrule(lr){12-16}
\# Agents & 2 & 3 & 4 & 5 & \emph{avg} & 2 & 3 & 4 & 5 & \emph{avg}
& 2 & 3 & 4 & 5 & \emph{avg} \\
\midrule
\multicolumn{16}{l}{\emph{Accuracy} (\%), HumanEval pass@1} \\
\dense{}                  & 74.5 & 75.2 & 75.2 & 75.2 & 75.0 & 82.7 & 86.0 & 84.7 & 84.7 & 84.5 & 87.5 & 86.2 & 85.6 & 85.0 & 86.1 \\
\kvcomm{}                 & 71.9 & 73.2 & 72.5 & 72.5 & 72.5 & 77.3 & 82.0 & 80.0 & 84.0 & 80.8 & 86.2 & 85.6 & 85.6 & 85.6 & 85.8 \\
CacheBlend, reuse-matched & 69.3 & 69.3 & 69.3 & 68.6 & 69.1 & 82.0 & \textbf{86.7} & 84.7 & 81.3 & 83.7 & 88.8 & 85.6 & 85.0 & 86.2 & 86.4 \\
CacheBlend, cost-matched  & 72.5 & 71.2 & 69.3 & 73.9 & 71.7 & \textbf{85.3} & 83.3 & 83.3 & 82.0 & 83.5 & 85.0 & 84.4 & 85.6 & 85.6 & 85.2 \\
\bcrchunk{}               & 74.5 & \textbf{73.9} & 72.5 & 75.2 & 74.0 & \cellcolor{tabhl}82.0 & \cellcolor{tabhl}86.0 & \cellcolor{tabhl}85.3 & \cellcolor{tabhl}\textbf{85.3} & \cellcolor{tabhl}\textbf{84.7} & 86.2 & 86.2 & 85.6 & 85.6 & 85.9 \\
\bcrgrow{}                & \cellcolor{tabhl}\textbf{77.1} & \cellcolor{tabhl}72.5 & \cellcolor{tabhl}\textbf{75.8} & \cellcolor{tabhl}\textbf{75.8} & \cellcolor{tabhl}\textbf{75.3} & 80.7 & 82.0 & 81.3 & 80.7 & 81.2 & \cellcolor{tabhl}\textbf{89.4} & \cellcolor{tabhl}\textbf{88.1} & \cellcolor{tabhl}\textbf{88.1} & \cellcolor{tabhl}\textbf{88.1} & \cellcolor{tabhl}\textbf{88.4} \\
\bcrsingle{}              & 71.2 & 71.9 & 73.9 & 73.2 & 72.5 & 84.0 & 83.3 & \textbf{86.0} & 84.7 & 84.5 & 88.8 & \textbf{88.1} & 86.9 & 86.9 & 87.7 \\
\midrule
\multicolumn{16}{l}{\emph{Reuse} (\%)} \\
\kvcomm{}                 & 45.1 & 47.4 & 47.3 & 46.9 & 46.7 & 64.9 & 60.0 & 55.2 & 52.1 & 58.1 & 89.8 & 87.8 & 85.6 & 84.1 & 86.8 \\
CacheBlend, reuse-matched & 46.0 & 49.5 & 48.5 & 48.9 & 48.2 & 65.6 & 60.3 & 55.6 & 52.2 & 58.4 & 87.9 & 87.2 & 85.4 & 83.9 & 86.1 \\
\bcrchunk{}               & 45.5 & 48.4 & 48.6 & 48.0 & 47.6 & 65.6 & 60.3 & 55.6 & 52.2 & 58.4 & 89.6 & 87.5 & 85.7 & 84.4 & 86.8 \\
\bcrgrow{}                & 46.4 & 50.8 & 52.3 & 52.0 & 50.4 & 63.8 & 58.8 & 53.9 & 50.6 & 56.8 & 89.2 & 87.8 & 86.2 & 84.9 & 87.0 \\
\bcrsingle{}              & 45.5 & 48.2 & 48.6 & 48.1 & 47.6 & 64.9 & 59.8 & 55.1 & 51.9 & 57.9 & 89.8 & 88.6 & 87.4 & 86.2 & 88.0 \\
\bottomrule
\end{tabular}
\end{table}

%% file: sections/allocation.tex
\subsection{Why a Budget, and at What Unit}
\label{sec:allocation}
\label{sec:alloc-calls}

\input{sections/table_evidence}

\paragraph{The allocation policies.}
One might instead reuse fewer caches and choose which. \Cref{tab:evidence}(a) runs the policies of App.~\ref{sec:method} on $150$ GSM8K problems with roles pinned: Protect-the-Root and Protect-the-Leaf force a dense prefill at the entry node and at the deepest node, and Random gives up reuse at calls drawn by a biased coin; \ogate{} is the exception, which \S\ref{sec:audit} measures in the reference configuration. The policies are distinguishable only by where they land on one line. Draw that line between \dense{} at reuse $0$ and unmodified \kvcomm{} at reuse $0.600$, scoring $85.3$ and $75.3$ in \Cref{tab:allocation}; its slope is $(85.3-75.3)/0.600=16.7$:
\begin{equation}
\text{accuracy}\ (\%) \;=\; 85.3 \;-\; 16.7 \times \text{reuse rate}.
\label{eq:line}
\end{equation}
The three policies between the endpoints land on the line, none more than $1.6$ points off (about half a standard error at $n{=}150$; the endpoints define the line, so only these three residuals are free). The sharpest case is Random, which consults nothing---not the anchors, not the answer, not the topology---yet reaches Protect-the-Root's reuse to within $0.002$ ($0.454$ against $0.456$) and carries the larger residual of the pair. The two structural policies land at reuse $0.456$ and $0.504$, and a divergence measurement says which should win: conditioned on predecessors that already matched \dense{}, the entry node diverges where the deepest does not, so the structural hypothesis predicts a gap. None resolves. The measurement was right about where error originates but wrong as a prescription: what determines the outcome is how much of the context was predicted, not where in the graph. One policy that fails to separate is weak evidence; four of different kinds---cache geometry, graph position twice, and nothing at all---make it a property of the problem. The cost is a function of how much you reuse, not of which calls. That claim is narrower than it sounds, for the reason the next measurement gives.

\paragraph{The unit of choice.}
\label{sec:granularity}
Large reported gains for choosing which \emph{tokens} to recompute sit awkwardly beside the null above for choosing which \emph{calls} to reuse. The two are not in conflict: the unit of choice is the variable, and choosing well loses value as the unit grows. To show this, we start from a reused cache of a $2048$-token segment, restore a nominal $20\%$ of its rows, and vary only how those rows are grouped and chosen---by attention from the consuming context's decode queries, or at random. Error is the relative $L_2$ distance of the attention output from a dense cache's, pooled over layers and heads because a recomputed span is shared by all of them. \Cref{tab:evidence}(b) gives the curve. Informed selection removes $49.5\%$ of the cache error beyond chance at single rows, $10.6\%$---a fifth of that---at $64$-token chunks, and $-5.7\%$ at $1024$, where it is nominally \emph{worse}; the prompts it wins fall from $6$ to $5$ to $1$ of six. \Cref{tab:granularity} carries the paired $t$ behind those counts ($p = 0.08$ at $64$) and notes that the budget drifts above the nominal $20\%$ once chunks are coarse, so $256$ and coarser are not budget-matched, pending a matched sweep. Two results that looked contradictory are thus one ordering, with our call-level null at one end and the literature's token-level gains at the other. The claim should be read with its unit attached: \emph{allocation does not matter across agent calls}. The mechanism is concentration. Within a segment the attention mass is extremely uneven, so there is much to choose between; across calls, the candidates are whole prompts of the same shape.

The chunk unit is where \kvcomm{}'s data structures can act, since its offset, anchor filter and re-rotation all assume a contiguous span, as App.~\ref{sec:method-chunk} details. The curve therefore predicts that at $64$-token chunks a selector keeps about a fifth of its single-row advantage, and that \repair{} should gain from repairing at all more than from repairing the right places. The task level bears out the ordering but not the fraction: in \Cref{tab:twobench}(a) the budget alone is worth $7.9$ points (the coin's $78.9$ over \kvcomm{}'s $71.0$) and the selector a further $5.4$ ($84.3$ over that coin), so the curve bounds the selector's value only loosely. Whether the curve's ordering by unit survives at the task level is the question of \Cref{tab:evidence}(c), which \S\ref{sec:grains} reads.

%% file: sections/table_evidence.tex
\begin{table}[!tbp]
\centering
\caption{\textbf{The evidence behind the cost, the two nulls and the layout ordering.} (a) five
agents, $150$ matched GSM8K problems, roles pinned; \emph{resid.} is the points earned beyond \Cref{eq:line}. (b) six prompts; \emph{gap}
is the cache error informed selection removes beyond chance, as a share of the unrepaired $0.1054$, \emph{won} the prompts where it beat the coin. (c) the three layouts of \S\ref{sec:bcr}, $b/c$ splits pooled over four agent counts. (d) is \Cref{tab:fairfull}'s cost rows, $p$ in parentheses. Throughout $b$ counts problems \kvcomm{} misses and the
other solves. Full tables in
\Cref{tab:allocation,tab:granularity,tab:fairfull}.}
\label{tab:evidence}
\renewcommand{\arraystretch}{0.9}

\begin{minipage}[t]{0.30\textwidth}
\centering\small\setlength{\tabcolsep}{3pt}
\begin{tabular*}{\linewidth}[t]{@{\extracolsep{\fill}}lcc}
\toprule
Policy & reuse & resid. \\
\midrule
Protect-the-Root     & 0.456 & $+0.9$ \\
Random           & 0.454 & $+1.6$ \\
Protect-the-Leaf & 0.504 & $+0.4$ \\
\midrule
\dense{}         & 0.000 & $+0.0$ \\
\kvcomm{}        & 0.600 & $+0.0$ \\
\bottomrule
\end{tabular*}

\vspace{3pt}
{\small (a) Allocation policies}
\end{minipage}\hfill
\begin{minipage}[t]{0.30\textwidth}
\centering\small\setlength{\tabcolsep}{3pt}
\begin{tabular*}{\linewidth}[t]{@{\extracolsep{\fill}}lcc}
\toprule
unit (rows) & gap \% & won of $6$ \\
\midrule
$1$     & $49.5\%$ & $6$ \\
$64$        & $10.6\%$ & $5$ \\
$1024$      & $-5.7\%$ & $1$ \\
\bottomrule
\end{tabular*}

\vspace{3pt}
{\small (b) Unit of choice}
\end{minipage}\hfill
\begin{minipage}[t]{0.34\textwidth}
\centering\small\setlength{\tabcolsep}{3pt}
\begin{tabular*}{\linewidth}[t]{@{\extracolsep{\fill}}lccc}
\toprule
 & chunk & grow & single \\
\midrule
GSM8K     & 47/24 & 41/39 & 44/22 \\
MMLU      & 21/12 & 54/37 & 50/50 \\
HumanEval & 4/3   & 29/12 & 27/15 \\
\bottomrule
\end{tabular*}

\vspace{3pt}
{\small (c) Layouts against \kvcomm{}}
\end{minipage}

\vspace{5pt}

\begin{minipage}[t]{0.85\textwidth}
\centering\footnotesize\setlength{\tabcolsep}{2pt}
\begin{tabular*}{\linewidth}[t]{@{\extracolsep{\fill}}lrrrrr}
\toprule
\dense{} vs.\ \kvcomm{} & 2 agents & 3 agents & 4 agents & 5 agents & pooled \\
\midrule
MMLU          & 16/12 (0.572) & 15/12 (0.701) & 18/14 (0.597) & 18/14 (0.597) & 67/52 \\
GSM8K         & 17/9 (0.169)  & 11/5 (0.210)  & 15/8 (0.210)  & 9/8 (1.000) & 52/30 \\
HumanEval     & 8/6 (0.791)   & 7/6 (1.000)   & 5/5 (1.000)   & 8/9 (1.000) & 28/26 \\
\bottomrule
\end{tabular*}

\vspace{3pt}
{\small (d) The grid's cells, $b/c$ ($p$), and their totals across the four counts}
\end{minipage}
\end{table}

%% file: sections/grains.tex
\subsection{What the Repair Recovers}
\label{sec:grains}

\dense{} prefills every segment and is the ceiling in the sense that matters: no budget $\rho<1$ can improve on $\rho{=}1$. \kvcomm{} is the method under study. CacheBlend \citep{yao2025cacheblend}, reproduced from its released rule inside this harness so that only its selection rule differs from \repair{}'s, is the published selective-recomputation competitor, and a coin at the same budget is the control. The gate is $\gate{=}0.3$ on the Llama benchmarks and $\gate{=}1.0$ on HumanEval; every repair takes $\rho{=}0.20$, with $\tau{=}0.25$ for \bcrgrow{}. Within each comparison the selection condition is therefore the only variable, with one exception: GSM8K's CacheBlend pairings run at $\gate{=}1.0$ against the other conditions' $0.3$, a mismatch that App.~\ref{app:cacheblend} shows runs against CacheBlend.

\input{sections/table_twobench}

\paragraph{The GSM8K recovery.} In the pinned configuration, \Cref{tab:twobench}(a) has the repair recovering $84.3 - 71.0 = 13.3$ of the $84.7 - 71.0 = 13.7$ points reuse costs, level with \dense{}. Against \kvcomm{} its discordant split is $9/49$ in that table's orientation ($9$ only \kvcomm{} solves, $49$ only the repair; $p<10^{-3}$). The run's reuse rate is $0.629$, at which \Cref{eq:line}, drawn through \Cref{tab:allocation}'s $n{=}150$ endpoints, predicts $85.3 - 16.7\times 0.629 = 74.8$; the repair sits $9.5$ points above the line that allocation could only slide along, because its budget is spent below the call rather than among calls. The direction repeats in \Cref{tab:fair}'s reference grid: \bcrchunk{} closes the gap in every GSM8K cell and is level with \dense{} at $26/25$ pooled, as App.~\ref{app:abovedense} shows. App.~\ref{app:withdrawn} withdraws a pre-registered pooled test; the $n{=}300$ run over distinct items carries this claim.

\paragraph{The random-selection control.} At the same budget we run a coin---the control the selective-recomputation literature competes without. One run scored $82.3\%$, and we first read its two-point gap to the draft-scored repair as a null. Five seeds withdraw that reading. Across seeds accuracy spans $74.3$ to $82.7\%$ over the five runs App.~\ref{app:negative} lists. The draft-scored repair is above all of them, three of the five paired tests separate the two, and \Cref{tab:twobench} reports the mean. On GSM8K at $64$-token chunks the selector is worth $84.3 - 78.9 = 5.4$ points, more than \S\ref{sec:granularity}'s curve predicted.

\paragraph{The layout ordering.} Every repair layout is tested against \kvcomm{} in every cell of \Cref{tab:fairfull}'s grid. Of the $36$ tests exactly one is significant---\bcrsingle{} on GSM8K at two agents, $14/4$, $p = 0.031$---and it does not survive App.~\ref{app:gridcell}'s correction for the comparisons it sits among. What the grid does give is direction, in the pooled splits of \Cref{tab:evidence}(c)---descriptive, because the four cells re-run the same problems. On GSM8K single rows lead \kvcomm{} and grown spans do not move; on MMLU the ordering inverts and weakens. HumanEval has no cost to recover: summing \Cref{tab:evidence}(d)'s $8/6$, $7/6$, $5/5$ and $8/9$ puts \dense{} at $28/26$ against \kvcomm{}, and no repair damages it.

\paragraph{The published selector.} \Cref{tab:fair} pairs CacheBlend at matched cost and at matched reuse, with \Cref{tab:budget}(b) counting the rows. Neither pairing separates from \bcrchunk{} in any cell of \Cref{tab:graincompare} ($p$ from $0.110$ to $1.000$), and \Cref{tab:twobench}(b) pools them. Ranking individual \emph{tokens} by a layer-1 probe buys nothing here over $64$-token chunks ranked by a two-token draft, though \S\ref{sec:granularity}'s curve predicts a gap. \Cref{tab:fair} also has \repair{} above \dense{} in sixteen of its thirty-six cells, and none of them shows repair beating exact computation: the splits are level, and tokenization makes a repaired call a \emph{different} input, free to land on either side of \dense{}, as App.~\ref{app:abovedense} explains.

\paragraph{The MMLU tokenization term.} MMLU is graded by letter matching, and four stricter extractors put its $43$-problem cost at $28$--$31$, so its accuracies are upper bounds. In the pinned configuration, \Cref{tab:twobench}(a) shows MMLU recovering nothing significant: $3.0$ points ($54.7$ against $51.7$) on $33$ discordant pairs ($12{+}21$), $p = 0.16$, while both reuse ($p<10^{-3}$) and repair sit far below \dense{}. The repair ran as designed---$1{,}253$ calls repairing $95{,}425$ rows, five skipped---and did not help. The reuse path concatenates per-segment IDs, and that sequence need not equal the chat template's tokenization of the same text. A token-matched control keeps the sequence but removes the approximation, by forcing every call to a dense prefill of the reuse path's merged IDs. It scores $59.3\%$, which splits the gap in two: $6.7$ points from tokenization ($66.0-59.3$) and $7.6$ from approximation ($59.3-51.7$; $25/48$, $p = 0.010$). \bcrchunk{} recovered $3.0$ of the $7.6$ and cannot touch the $6.7$, because those points are not in the cache but in which tokens the cache is \emph{of}. On GSM8K the entry tokenizations coincide, the term vanishes, and repair recovers nearly the whole gap---the benchmark asymmetry in one line. Any reuse system that concatenates per-segment IDs pays this tax.

\paragraph{What the repair buys.} At a budget of a fifth of the cache, \repair{} returns GSM8K to \dense{} while $63\%$ of calls are still served from cache, and by \Cref{tab:budget}(c) reaches the first token $2.85\times$ sooner than \dense{} does. In \Cref{tab:fair} a \repair{} layout holds the best mean short of \dense{} in thirteen of fifteen columns and in every \emph{avg} column, and the best layout's average is at or above \dense{}'s on all three benchmarks. CacheBlend leads no average at either pairing, and its reuse-matched pairing recomputes more than twice \bcrchunk{}'s rows per call. The draft is what pays for this: at the same budget it beats a coin by $5.4$ points, and no layout damages the benchmark that has nothing to recover. Where it recovers little, on MMLU, almost half of the gap is one no budget could reach, because it is not in the cache.

%% file: sections/table_twobench.tex
\begin{table}[!t]
\centering
\caption{\textbf{Where repair works, and what it is worth against a published selector.}
Five agents pinned to one role, $300$ matched problems. In (a) $b$ counts problems
\kvcomm{} solves at its shipped $\gate{=}0.3$ and the row misses, $c$ the reverse, $p$ in
parentheses, $^{\ast}$ marking $p<0.05$; $-$ marks a row with
no split of its own---\kvcomm{} itself, the $^{\dagger}$ five-seed coin mean,
and the token-matched control, which only MMLU needs. Shading marks the repair row. (b) pools the reference grid over four agent
counts, descriptively
(\Cref{tab:graincompare}).}
\label{tab:twobench}

\begin{minipage}[t]{0.66\textwidth}
\centering\footnotesize\setlength{\tabcolsep}{2.5pt}
\begin{tabular*}{\linewidth}{@{\extracolsep{\fill}}lcr@{\,}lcr@{\,}l}
\toprule
& \multicolumn{3}{c}{GSM8K} & \multicolumn{3}{c}{MMLU} \\
\cmidrule(lr){2-4}\cmidrule(lr){5-7}
Method & acc. & \multicolumn{2}{c}{$b/c$ ($p$)} & acc. & \multicolumn{2}{c}{$b/c$ ($p$)} \\
\midrule
\kvcomm{}              & 71.0                    & \multicolumn{2}{c}{--} & 51.7 & \multicolumn{2}{c}{--} \\
\bcrchunk{}, coin      & 78.9\rlap{$^{\dagger}$} & \multicolumn{2}{c}{--} & 50.0\rlap{$^{\dagger}$} & \multicolumn{2}{c}{--} \\
\hi \bcrchunk{}, draft & 84.3                    &  9/49 & ($0.000^{\ast}$) & 54.7 & 12/21 & ($0.163$)        \\
token-matched \dense{} & --                      & \multicolumn{2}{c}{--} & 59.3 & 25/48 & ($0.010^{\ast}$) \\
\dense{}               & 84.7                    & 10/51 & ($0.000^{\ast}$) & 66.0 & 21/64 & ($0.000^{\ast}$) \\
\bottomrule
\end{tabular*}

\vspace{3pt}
{\small (a) Cost and repair at $n{=}300$}
\end{minipage}\hfill
\begin{minipage}[t]{0.32\textwidth}
\centering\footnotesize\setlength{\tabcolsep}{2.5pt}
\begin{tabular*}{\linewidth}{@{\extracolsep{\fill}}lccc}
\toprule
pairing & GSM8K & MMLU & HEval \\
\midrule
reuse & 25/31 & 39/69 & 29/26 \\
cost  & 23/30 & 47/61 & 21/26 \\
\bottomrule
\end{tabular*}

\vspace{3pt}
{\small (b) CacheBlend vs.\ \bcrchunk{}}
\end{minipage}
\end{table}

%% file: sections/conclusion.tex
\section{Discussion and Conclusion}
\label{sec:conclusion}
\label{sec:discussion}

Cross-context KV-cache reuse is reported to be lossless. On two of three benchmarks we find otherwise. With roles pinned, reuse costs $13.7$ points on GSM8K and $14.3$ on MMLU ($9.3$--$10.3$ after re-grading), while HumanEval shows no measurable loss. Nor can it be avoided by choosing which calls to reuse: the value of informed selection falls from $49.5\%$ to $10.6\%$ to $0$ as the unit of decision grows. It can be repaired: a fixed budget of exact recomputation lets \repair{} restore GSM8K to \dense{} accuracy while keeping most of the reuse (reuse rate $0.629$), and two of the three layouts reach that level; on every benchmark of the reference grid the best layout leads every reuse baseline's mean. On MMLU, however, only part of the gap is in the cache at all.

The key finding is that \emph{where} the decision is made matters as much as \emph{whether} it is. Our evidence spans three benchmarks and two backbones, but a single topology and a single chunk size. \repair{} also trades latency for accuracy: its $2.85\times$ speedup in \Cref{tab:budget}(c) is below reuse's $7.63\times$, a trade App.~\ref{app:limitations} weighs. Testing the grid's ordering cell by cell, carrying the protocol to other reuse systems, and closing that latency gap are the work the same appendix lays out.

%% file: sections/statements.tex

\subsection*{AI use statement}
We used generative AI tools solely for language editing, including grammar correction and concise rewriting. All research, ideas, and manuscript content were developed by the authors. We reviewed all AI-assisted edits and take full responsibility for the final content.

%% file: sections/method.tex
\section{Allocation Policies in Detail}
\label{sec:method}

The policies we test were not chosen to be exhaustive but to be \emph{different
in kind}, so that a common outcome cannot be blamed on a common blind spot.
They are organised by the unit they allocate over.

\subsection{Allocating whole agent calls}

\input{sections/table_warmblocks}
\input{sections/table_allocation}

\paragraph{P1: the entropy gate (what \kvcomm{} ships).}
The baseline policy is \Cref{eq:gate}, swept from the shipped $\gate{=}0.3$ to
$1.0$.

\paragraph{P2: \ogate{}, a gate on the answer's opening.}
Eviction's fix for a miscalibrated prompt-side score is to defer the decision
past the first generated tokens. \ogate{} transplants that fix: serve
optimistically, generate $k$ tokens on the assembled cache, decide from the
confidence of those $k$ distributions, and pay a dense prefill when the draft is
not trusted.

The literal transplant does not carry over, and it is worth saying why. In
eviction the exact cache exists for free at the end of prefill, so deferring
costs only $k$ decode steps; in reuse the exact cache is \emph{never
computed}, and obtaining it costs the prefill that reuse exists to avoid.
What can be moved is the \emph{decision}, not the approximation.

\paragraph{P3 and P4: structural allocation.}
A third kind decides from the shape of the request rather than from a signal,
before it starts.
\emph{Protect-the-Root} forces a dense prefill at the entry node;
\emph{Protect-the-Leaf} does the same at the deepest node. They force the
identical number of dense calls and differ only in \emph{which} node is
protected, which makes them a matched pair.

\paragraph{P5: random allocation.}
Give up a fixed fraction of reuse opportunities by biased coin flip, with the
probability chosen so the achieved reuse rate lands on top of the structural
policies.

\subsection{Allocating below the call: \bcrchunk{}}
\label{sec:method-chunk}

\input{sections/table_granularity}

\paragraph{Chunks rather than tokens.}
The replacement policies above are all-or-nothing per agent call, at or above
the unit of the per-segment gate they replace, and \S\ref{sec:granularity}'s
curve points at a finer one. Per-row selection is where the choice is worth
most, and it is what the
selective-recomputation literature for RAG does. It is also what \kvcomm{}'s
data structures cannot express. The offset of \Cref{eq:delta} is a
position-indexed elementwise tensor, anchors are filtered by a length
comparison, and re-rotation applies one scalar positional shift to a segment;
all three assume the reused unit is a dense, contiguous, length-typed span.
A contiguous chunk satisfies every one of those assumptions, and recomputing
one is an ordinary prefill rather than a custom attention path.

\paragraph{The policy.}
On every call \kvcomm{} serves from an assembled cache, \repair{} inserts three
steps before generation. A prompt-side score would be free at that moment---\kvcomm{}'s
own gate is one such score---but any such score ranks cached rows by the geometry
of the cache rather than by what the answer turns out to need. We therefore spend
$k$ decode steps on a draft instead.
\begin{enumerate}[leftmargin=*]
\small
\item \textbf{Draft.} Generate $k{=}2$ tokens on the assembled cache and keep
the last $k$ query rows. The draft is then discarded: every layer is truncated
back to its pre-draft length and the token counter restored. Repair and real
generation therefore both begin from the cache the call was served with. The $k$
drafted tokens are decoded a second time, now against the repaired cache, and
need not come back the same. \ogate{} handles its draft the other way: it keeps the
draft and continues from the returned cache (App.~\ref{app:ogate}). That is the
case in which ``no extra forward pass'' holds; for \repair{} it does not. The served cache
stops one token short of the prompt ($T = |x| - 1$), so at $k{=}2$ the kept rows
are the queries of the prompt's final token and of the first generated one: one
prompt-side, one answer-side. We fix $k{=}2$ and never sweep it, so
\S\ref{sec:grains}'s estimate bounds what one answer-side query row is worth,
not what a longer draft would be.
\item \textbf{Score.} The rule of \S\ref{sec:bcr}, with grouped-query keys
expanded to query heads. Scoring a chunk by the mean of its rows' scores keeps
a short trailing chunk from being penalised for its length, and unlike an
eviction mask the selection cannot differ per head.
\item \textbf{Repair.} Recompute the highest-scoring $\rho$ of the chunks
exactly against the cache, and resume.
\end{enumerate}

The skip rule, the recompute path, the budget's denominator and the correctness
check are stated once in App.~\ref{app:selective}, with
\Cref{alg:restitch}.

\paragraph{The random control.}
Repairing chunks and \emph{choosing} them are separable, and \S\ref{sec:granularity}
predicts the choice is worth much less at the chunk unit than at the row
unit. We therefore run the same policy with the chunks picked by coin
at an identical budget. Whatever separates the two rows is the draft's
contribution; whatever they share belongs to repair itself. The coin condition also
skips the draft entirely, so it spends $k$ fewer decode steps per repaired call:
a null between the rows would have been a dominance rather than a tie. A single
coin run had suggested such a null; the five-seed control of \S\ref{sec:grains}
replaces it, so the draft is not free to remove, and App.~\ref{app:latency}
prices what the timed rows omit.

%% file: sections/table_warmblocks.tex
\begin{table}[!t]
\centering
\caption{\textbf{Accuracy and reuse rate by block.} \dense{} against the shipped default
($\gate{=}0.3$) in blocks of $60$ GSM8K problems, with that block's reuse rate.
The shaded first block---what a $60$-problem evaluation would see---is the only
one where reuse does not lose, and the one with the lowest reuse rate.}
\label{tab:warmblocks}
\small
\begin{tabular}{lcccc}
\toprule
Problems & reuse rate & \dense{} & \kvcomm{} $\gate{=}0.3$ & $\Delta$ \\
\midrule
\hi 1--60 & 0.537 & 80.0 & 81.7 & $+1.7$ \\
61--120 & 0.625 & 88.3 & 80.0 & $-8.3$ \\
121--180 & 0.679 & 90.0 & 85.0 & $-5.0$ \\
181--240 & 0.675 & 83.3 & 80.0 & $-3.3$ \\
241--300 & 0.675 & 86.7 & 80.0 & $-6.7$ \\
\midrule
1--300 & 0.638 & 85.7 & 81.3 & $-4.3$ \\
\bottomrule
\end{tabular}
\end{table}

%% file: sections/table_allocation.tex
\begin{table}[!t]
\centering
\caption{\textbf{Allocation policies on one line.} Five agents, $150$ matched GSM8K
problems, roles pinned; rows sorted by reuse rate. \emph{pred.} is what
\Cref{eq:line} predicts from the reuse rate alone, \emph{resid.} the points
earned beyond it. $p$ is an exact McNemar test against \dense{}; $^{\ast}$ marks
$p<0.05$. Both columns use the unrounded scores, so redoing them from
the printed ones can shift the last digit.}
\label{tab:allocation}
\small
\begin{tabular}{llccccrr}
\toprule
Policy & decides from & reuse & acc.\ (\%) & 95\% CI & $p$ vs \dense{} & pred. & resid. \\
\midrule
\dense{} & no reuse & 0.000 & 85.3 & {\scriptsize [79, 90]} & -- & 85.3 & $+0.0$ \\
Protect-the-Root & structural & 0.456 & 78.7 & {\scriptsize [71, 84]} & 0.013\rlap{$^{\ast}$} & 77.7 & $+0.9$ \\
Random & information-free & 0.454 & 79.3 & {\scriptsize [72, 85]} & 0.064 & 77.8 & $+1.6$ \\
Protect-the-Leaf & structural & 0.504 & 77.3 & {\scriptsize [70, 83]} & 0.023\rlap{$^{\ast}$} & 76.9 & $+0.4$ \\
\kvcomm{} $\gate{=}0.3$ & entropy gate & 0.600 & 75.3 & {\scriptsize [68, 82]} & 0.006\rlap{$^{\ast}$} & 75.3 & $+0.0$ \\
\bottomrule
\end{tabular}
\end{table}

%% file: sections/table_granularity.tex
\begin{table}[!t]
\centering
\caption{\textbf{Selection value by unit of choice.} One $2048$-token segment, six
prompts, varying only how rows are grouped and picked. Error is the relative
$L_2$ distance of the attention output from a dense cache ($0.1054$ unrestored);
\emph{gap} is what informed selection removes beyond chance; the last column
gives prompts where informed won and a paired $t$ over the six. The budget keeps
$\mathrm{max}(1, \mathrm{round}(0.2 n_{\mathrm{chunks}}))$ chunks, so the last
three rows are not budget-matched.}
\label{tab:granularity}
\small
\setlength{\tabcolsep}{4pt}
\begin{tabular}{llccccc r@{\,}l}
\toprule
unit of choice & budget & attention & random & gap & gap / no-repair
& \multicolumn{2}{c}{prompts (\emph{p})} \\
\midrule
$1$ (single row) & $20.0\%$ & $0.0375$ & $0.0896$ & $+0.0521$ & $49.5\%$ & $6/6$ & ($0.01$)\\
$16$ & $20.3\%$ & $0.0650$ & $0.0851$ & $+0.0201$ & $19.0\%$ & $6/6$ & ($0.02$)\\
$64$ & $18.8\%$ & $0.0763$ & $0.0875$ & $+0.0112$ & $10.6\%$ & $5/6$ & ($0.08$)\\
$256$ & $25.0\%$ & $0.0780$ & $0.0880$ & $+0.0100$ & $9.5\%$ & $4/6$ & ($0.28$)\\
$1024$ & $50.0\%$ & $0.0691$ & $0.0631$ & $-0.0060$ & $-5.7\%$ & $1/6$ & ($0.69$)\\
$2048$ (whole segment) & $100.0\%$ & \multicolumn{2}{c}{\emph{both restore everything}} & $0$ & $0\%$ & \multicolumn{2}{c}{--} \\
\bottomrule
\end{tabular}
\end{table}

%% file: sections/selective.tex
\section{Selective Recomputation}
\label{app:selective}

\begin{algorithm}[t]
\caption{\repair{} (Budgeted Cache Repair): repair a fixed fraction of an assembled cache, then decode.}
\label{alg:restitch}
\footnotesize
\begin{algorithmic}[1]
\REQUIRE prompt $x_{1:M+1}$; assembled cache
$\widetilde{K}^{\ell},\widetilde{V}^{\ell}$ over the $M$ cached rows
(\Cref{eq:delta}; the reuse path slices the cache one token short of the
prompt); chunk size $B{=}64$; budget $\rho{=}0.2$
\STATE \textbf{if} $M < 2B$ \textbf{or} $\rho M < B$ \textbf{then} serve unrepaired
\STATE partition $1{:}M$ into $B$-token chunks; $\mathcal{R} \gets$ rows of the
$\operatorname{round}(\rho M/B)$ highest-scoring ones
\COMMENT{\Cref{alg:draft}; better than a coin (\Cref{tab:twobench})}
\STATE $h \gets \operatorname{embed}(x_{\mathcal{R}})$
\FOR{layer $\ell = 1,\dots,L$}
  \STATE $q,k,v \gets$ layer-$\ell$ projections of $h$, rotary-encoded at
  $\mathcal{R}$'s true positions
  \STATE $\widetilde{K}^{\ell}[\mathcal{R}],\widetilde{V}^{\ell}[\mathcal{R}] \gets k,v$
  \COMMENT{overwrite \emph{before} attending}
  \STATE $h \gets$ block $\ell$: causal attention of $q$ over
  $\widetilde{K}^{\ell},\widetilde{V}^{\ell}$ up to $\max P$, then MLP
\ENDFOR
\ENSURE rows in $\mathcal{R}$ exact, the rest assembled; decoding unchanged
\end{algorithmic}
\end{algorithm}

\begin{algorithm}[t]
\caption{Draft scoring: rank cached positions by what the answer's opening attends to.}
\label{alg:draft}
\footnotesize
\begin{algorithmic}[1]
\REQUIRE assembled cache $\widetilde{K}^{\ell},\widetilde{V}^{\ell}$ of length $M$; prompt ids; draft length $k{=}2$
\STATE $\Lambda \gets$ per-layer cache lengths; $c \gets$ cache token counter
\STATE generate $k$ tokens greedily on the cache, recording each layer's query projections
\STATE truncate $\widetilde{K}^{\ell},\widetilde{V}^{\ell}$ to $\Lambda$; restore $c$
\COMMENT{the draft is discarded, not continued}
\STATE $Q^{\ell} \gets$ the last $k$ recorded query rows at layer $\ell$
\COMMENT{the prompt's final token, then the first $k-1$ generated ones}
\STATE score positions and chunks by the draft rule of \S\ref{sec:bcr}
\COMMENT{grouped-query keys expanded to query heads}
\ENSURE chunk $i$ scored by $\operatorname{mean}_{j \in i} a_j$
\end{algorithmic}
\end{algorithm}

\paragraph{The recompute path.}
Repairing a chunk exactly requires everything before it, so recomputing it in
isolation is wrong and recomputing the prefix again saves nothing. We run only
the selected positions forward while they attend to the cache already present.
Each layer does four things. The selected rows' hidden states go through the
block's projections and rotary embedding. Their new keys and values overwrite
their own slots \emph{before} attention is taken, so a repaired row attending to
another repaired row sees the corrected value. Their queries then attend over the
whole cache under a causal mask built from absolute positions. The block output
feeds the next layer. Layers proceed in order, so later layers read the
corrections earlier ones made. The cost is
$O(\text{selected}\times\text{total})$ attention rather than
$O(\text{total}^2)$.

\paragraph{Correctness.}
Selecting every position must reproduce \dense{}. Compared tensor by tensor,
the two agree up to numerical precision: layer 0 is exact, and the
relative difference grows with depth to a few percent at 512 tokens, because this path
accumulates attention in float32 while the model's fused kernel does not. The functional
check is the one that decides, and it passes---greedy generation from the two
caches is token-identical over 24 tokens. We report the tensor-level
discrepancy rather than tune it away.

\paragraph{The seam, and what the budget counts.}
We install the policy by wrapping \texttt{model.generate}, the same seam
\ogate{} uses, and for the same reason: every state mutation in \kvcomm{}'s reuse
path happens after generate returns. We rank chunks over the whole cached prefix rather than over the reused segment
alone, because segment boundaries are not visible at that seam. Budget spent on a
chunk that was already exact therefore has no effect, and the row counts we
report include it.

%% file: sections/experiments.tex
\section{Latency, and What the Repair Costs}
\label{app:latency}

\input{sections/table_gamma}
\input{sections/table_breakeven}
\input{sections/table_rho}
\input{sections/table_maxspan}
\input{sections/table_fairfull}

Latencies come from $n{=}60$ runs with one process on an idle GPU
(\Cref{tab:gamma}); the larger runs were executed sequentially, and we report
no headline latencies from them---the one per-problem timing taken from an
$n{=}150$ run is reported in App.~\ref{app:negative}, where it does not pass a
paired test.

\paragraph{Latency on the workload.} Against \dense{}'s $101.5$\,ms, every reuse setting lands between
$78.8$ and $82.7$\,ms---a $1.2$--$1.3\times$ improvement, flat in $\gate$
because the calls converted at high $\gate$ are the ambiguous ones whose
cache assembly costs nearly what it replaces. The prefill being replaced is
small: GSM8K questions average $59$ tokens against roughly $1{,}000$
generated across four agent calls, so prefill is $\approx 0.4\%$ of a
request and the $22$\,ms recovered is a fifth of it. End-to-end,
latency rises rather than falls, and rises \emph{with} reuse---$28.33$\,s \dense{}
against $32.00$\,s at the default gate, rising to $36.78$\,s---which puts
the per-call cost of anchor matching and offset application near
$0.9$--$2.1$\,s against the $22$\,ms of prefill it removes.

\paragraph{Scaling with segment length.}
\Cref{tab:breakeven} repeats the measurement on synthetic prompts of
controlled length. \kvcomm{}'s time-to-first-token is nearly flat in the
segment---$56$ to $73$\,ms across a $64\times$ range---while a dense prefill
of the same prompt grows from $159$ to $543$\,ms, so the ratio grows with
the segment to $7.6\times$ at $4096$ symbols, reproducing this method's
headline figure. The advantage is structural: assembling a cache from
anchors does not scale with the segment the way prefilling it does.

\paragraph{The cost of repair.}
The \repair{} rows repeat that sweep with \bcrchunk{} installed in its
minimal form---random selection, no draft, the chunk attention fused; the
FA2 rows route each chunk through FlashAttention-2's bottom-right-aligned
causal kernel, which needs neither mask nor query folding. The premium over
plain reuse is the price of the repaired rows and is linear in $\rho$, as
the arithmetic requires: at $4096$ symbols the repair costs $46$, $64$,
$117$, $182$ and $231$\,ms at $\rho = 0.05$ through $0.40$, for speedups of
$4.5\times$ down to $1.8\times$.

\paragraph{The accuracy--latency frontier.}
Reuse reaches the first token $7.6\times$ sooner and pays
\Cref{tab:fair}'s accuracy loss, while \repair{} reaches it $2.9\times$
sooner at $\rho{=}0.20$ and, by the same table, pays none of it on GSM8K or
HumanEval; on MMLU that offline grid separates no condition from \dense{}
either way (\Cref{tab:fairfull} for reuse, App.~\ref{app:abovedense} for the
repair), while the pinned run of \S\ref{sec:grains} puts both far below it.
For the coin-selected repair, \Cref{tab:rho} shows the budget buying latency
but no accuracy.

\paragraph{What the sweep does not show.}
Two caveats. The end-to-end columns move by $0.6$--$3.3\%$ for plain reuse and
by under $6\%$ for any repair row, buried under a generation identical in
every condition, so they measure no end-to-end effect in either direction. And
the sweep runs 12 samples per length, so the anchor pool never
saturates---while the per-call overhead that makes reuse slower end-to-end on
GSM8K grows with the number of anchors blended. \Cref{tab:breakeven} is
therefore the best case for reuse, not its steady state.

\paragraph{Placement at a fixed budget.} Everything so
far varies \emph{how much} cache is recomputed. \Cref{tab:maxspan} varies
only \emph{how it is arranged}: the budget stays $\mathrm{round}(\rho M)$
and the cap on one contiguous repaired span moves from a single token up to
$64$. The result is a step rather than a slope. Repairing individual rows scores above \dense{} in both cells; allowing spans to
grow at all costs most of that immediately, and growing them further changes
nothing---the $8$-, $16$- and $64$-token caps land within one problem of
each other, and their realised mean span lengths sit within $0.2$ tokens
because $\tau$, not the cap, is what stops growth. Pooled over all four
agent counts (the two- and four-agent cells are those of \Cref{tab:maxspan}; the other two are not tabulated), \bcrsingle{} leads the $64$-token cap $47/27$; the naive McNemar
$p$ is $0.027$, but the four cells re-run the same problems
(\S\ref{sec:setup}) and the item-level correction returns $p = 0.13$.

The table also separates the unit from the rule that picks it. CacheBlend at
the matched cost is a single-row layout too, ranking rows by their layer-1
value deviation instead of by our draft score, and it lands with the $1$-token
cap---one cell each way, neither significant. Two unrelated selectors put the
same budget on single rows and land together, which is
\S\ref{sec:granularity}'s conclusion reached in task space.

Two qualifications. First, the comparison the body's claims rest on is against \bcrchunk{},
and there the two are indistinguishable ($25/26$ pooled), exactly as both are
indistinguishable from \dense{}: \bcrsingle{} is an equivalent way to spend the
budget on this workload rather than a better one, and its advantage over \bcrgrow{}
does not carry to MMLU. Second, the ordering across the sweep is the evidence rather than any single
contrast (\S\ref{sec:grains})---read with \Cref{tab:rho}'s caution about
single-run orderings, though the two settings whose caps do not bind
returned identical answers on all $150$ problems, which bounds this scale's
run-to-run noise at zero.

\paragraph{Implementations.}
The accuracy tables in the body were produced with the earlier draft-scored,
layer-serial repair; the timing rows use the coin-selected, fused one. The two
are functionally equivalent---repairing every chunk reproduces \dense{}'s
greedy generation token-for-token under both---but not equal in cost: the
draft-scored repair spends $k{=}2$ decode steps per repaired call before any repair
begins, and \Cref{tab:breakeven} does not charge for them. Since the control of
\S\ref{sec:grains} shows the draft contributes to accuracy, the timed rows are
to be read as the cost of the coin condition; the draft-scored configuration
adds the $k$ decode steps on top of them.

%% file: sections/table_gamma.tex
\begin{table}[!t]
\centering
\caption{\textbf{The gate sweep at $n{=}60$.} The threshold $\gate$ over its meaningful
range on $60$ matched GSM8K problems; $\gate{=}1.0$ disables the entropy test, so
the last row is unconditional reuse. \emph{all}/\emph{late} are the fraction of
calls served without a dense prefill, over all $60$ problems and the last $20$;
\emph{calls} counts reused versus dense calls out of $240$. $p$ is an exact
McNemar against the $\gate{=}0.3$ default; brackets are Wilson
$95\%$ intervals.}
\label{tab:gamma}
\small
\begin{tabular}{llcccccrrr}
\toprule
& & \multicolumn{3}{c}{Accuracy} & \multicolumn{3}{c}{Reuse} &
\multicolumn{2}{c}{Latency} \\
\cmidrule(lr){3-5}\cmidrule(lr){6-8}\cmidrule(lr){9-10}
Method & $\gate$ & acc.\ (\%) & 95\% CI & $p$ & all & late & calls &
TTFT (ms) & E2E (s) \\
\midrule
\dense{} & -- & 80.0 & {\scriptsize [68, 88]} & -- & 0.000 & 0.000 & 0/240 & 101.5 & 28.33 \\
\kvcomm{} & $0.3$ & 81.7 & {\scriptsize [70, 89]} & -- & 0.537 & 0.662 & 129/111 & 79.2 & 32.00 \\
\kvcomm{} & $0.5$ & 81.7 & {\scriptsize [70, 89]} & 1.00 & 0.671 & 0.850 & 161/79 & 82.7 & 36.78 \\
\kvcomm{} & $0.7$ & 78.3 & {\scriptsize [66, 87]} & 0.73 & 0.787 & 0.875 & 189/51 & 79.6 & 36.70 \\
\kvcomm{} & $1.0$ & 78.3 & {\scriptsize [66, 87]} & 0.62 & 0.838 & 0.950 & 201/39 & 78.8 & 36.33 \\
\bottomrule
\end{tabular}
\end{table}

%% file: sections/table_breakeven.tex
\begin{table}[!t]
\centering
\caption{\textbf{Latency against segment length.} Five \texttt{CopyMachine} agents on
synthetic prompts. The harness runs each call twice in one process---assembled
cache and full prefill---so every ratio is paired, and the speedup rows are
means of those per-call ratios rather than ratios of the means above them; the
two differ, $7.63\times$ against $542.7/72.9$. $\rho$ is the repaired
fraction, $0$ plain reuse and $1$ a full prefill; the parenthetical names the
repair kernel: PyTorch's scaled dot-product attention, or FlashAttention-2. \emph{Caveat:} $12$ samples per length, so the anchor pool never
saturates---the best case for reuse, not its steady state. Shading marks the $\rho{=}0.20$ rows, the setting used elsewhere.}
\label{tab:breakeven}
\footnotesize
\setlength{\tabcolsep}{5pt}\renewcommand{\arraystretch}{0.95}
\begin{tabular}{llccccc}
\toprule
& & & \multicolumn{4}{c}{Shared segment (symbols)} \\
Metric & Method & $\rho$ & 64 & 512 & 2048 & 4096 \\
\midrule
\multirow{8}{*}{TTFT (ms)} & \dense{} & 1.00 & 159.2 & 198.1 & 341.2 & 542.7 \\
 & \kvcomm{} & 0.00 & 60.3 & 56.2 & 64.7 & 72.9 \\
 & \cellcolor{tabhl}\bcrchunk{} (SDPA) & \cellcolor{tabhl}0.20 & \cellcolor{tabhl}102.4 & \cellcolor{tabhl}112.1 & \cellcolor{tabhl}155.2 & \cellcolor{tabhl}229.8 \\
 & \cellcolor{tabhl}\textbf{\bcrchunk{} (FA2)} & \cellcolor{tabhl}\textbf{0.20} & \cellcolor{tabhl}\textbf{101.2} & \cellcolor{tabhl}\textbf{109.8} & \cellcolor{tabhl}\textbf{140.5} & \cellcolor{tabhl}\textbf{190.1} \\
 & \bcrchunk{} (FA2) & 0.10 & 100.8 & 99.0 & 111.1 & 137.0 \\
 & \bcrchunk{} (FA2) & 0.05 & 89.0 & 100.3 & 108.1 & 119.2 \\
 & \bcrchunk{} (FA2) & 0.30 & 113.8 & 124.6 & 173.4 & 255.1 \\
 & \bcrchunk{} (FA2) & 0.40 & 126.1 & 144.8 & 212.8 & 304.3 \\
\cmidrule(lr){1-7}
\multirow{7}{*}{Speedup vs.\ \dense{}} & \kvcomm{} & 0.00 & 2.48$\times$ & 3.26$\times$ & 5.25$\times$ & 7.63$\times$ \\
 & \cellcolor{tabhl}\bcrchunk{} (SDPA) & \cellcolor{tabhl}0.20 & \cellcolor{tabhl}1.56$\times$ & \cellcolor{tabhl}1.77$\times$ & \cellcolor{tabhl}2.20$\times$ & \cellcolor{tabhl}2.35$\times$ \\
 & \cellcolor{tabhl}\textbf{\bcrchunk{} (FA2)} & \cellcolor{tabhl}\textbf{0.20} & \cellcolor{tabhl}\textbf{1.58$\times$} & \cellcolor{tabhl}\textbf{1.81$\times$} & \cellcolor{tabhl}\textbf{2.43$\times$} & \cellcolor{tabhl}\textbf{2.85$\times$} \\
 & \bcrchunk{} (FA2) & 0.10 & 1.59$\times$ & 2.00$\times$ & 3.08$\times$ & 3.96$\times$ \\
 & \bcrchunk{} (FA2) & 0.05 & 1.80$\times$ & 1.97$\times$ & 3.16$\times$ & 4.54$\times$ \\
 & \bcrchunk{} (FA2) & 0.30 & 1.41$\times$ & 1.59$\times$ & 1.97$\times$ & 2.13$\times$ \\
 & \bcrchunk{} (FA2) & 0.40 & 1.27$\times$ & 1.37$\times$ & 1.61$\times$ & 1.78$\times$ \\
\cmidrule(lr){1-7}
\multirow{8}{*}{End-to-end (s)} & \dense{} & 1.00 & 14.95 & 14.11 & 14.74 & 15.06 \\
 & \kvcomm{} & 0.00 & 14.86 & 13.96 & 14.47 & 14.57 \\
 & \cellcolor{tabhl}\bcrchunk{} (SDPA) & \cellcolor{tabhl}0.20 & \cellcolor{tabhl}14.41 & \cellcolor{tabhl}14.42 & \cellcolor{tabhl}13.95 & \cellcolor{tabhl}14.54 \\
 & \cellcolor{tabhl}\textbf{\bcrchunk{} (FA2)} & \cellcolor{tabhl}\textbf{0.20} & \cellcolor{tabhl}\textbf{14.10} & \cellcolor{tabhl}\textbf{14.22} & \cellcolor{tabhl}\textbf{14.26} & \cellcolor{tabhl}\textbf{14.64} \\
 & \bcrchunk{} (FA2) & 0.10 & 15.03 & 14.22 & 14.31 & 14.48 \\
 & \bcrchunk{} (FA2) & 0.05 & 13.52 & 15.46 & 14.96 & 14.97 \\
 & \bcrchunk{} (FA2) & 0.30 & 14.22 & 14.48 & 14.24 & 14.38 \\
 & \bcrchunk{} (FA2) & 0.40 & 14.23 & 14.23 & 14.42 & 14.11 \\
\bottomrule
\end{tabular}
\end{table}

%% file: sections/table_rho.tex
\begin{table}[!t]
\centering
\caption{\textbf{The repair budget swept on GSM8K, at four coin seeds.} Two agents, $150$
matched problems, $64$-token chunks drawn by coin so that $\rho$ is the only
variable; \kvcomm{} scores $77.3\%$, \dense{} $82.7\%$. \emph{Each budget is run
four times}: the coin's stream advances per repaired call, so one run is one
realisation of the policy, not the policy.}
\label{tab:rho}
\small
\setlength{\tabcolsep}{5pt}\renewcommand{\arraystretch}{0.95}
\begin{tabular}{lcccccc}
\toprule
& \multicolumn{4}{c}{Accuracy (\%), by coin seed} & &  \\
\cmidrule(lr){2-5}
$\rho$ & 0 & 1 & 2 & 3 & mean & 95\% CI \\
\midrule
0.05 & 78.0 & 74.7 & 78.0 & 75.3 & 76.5 & $\pm$2.8 \\
0.10 & 76.7 & 76.0 & 82.0 & 75.3 & 77.5 & $\pm$4.9 \\
0.20 & 70.0 & 78.0 & 80.7 & 81.3 & 77.5 & $\pm$8.3 \\
0.30 & 81.3 & 78.0 & 75.3 & 82.7 & 79.3 & $\pm$5.3 \\
0.40 & 78.7 & 77.3 & 79.3 & 80.0 & 78.8 & $\pm$1.8 \\
\midrule
\kvcomm{} ($\rho{=}0$) & \multicolumn{4}{c}{77.3} & 77.3 & -- \\
\dense{} ($\rho{=}1$) & \multicolumn{4}{c}{82.7} & 82.7 & -- \\
\bottomrule
\end{tabular}
\end{table}

%% file: sections/table_maxspan.tex
\begin{table}[!t]
\centering
\caption{\textbf{The repair's layout, swept with its budget held fixed.} GSM8K,
$n{=}150$, $\rho{=}0.20$, draft-scored, two and four agents; the foot rows
reproduce \Cref{tab:fair}'s GSM8K columns. Only the cap on one contiguous span
moves. Cells not run are dashed; $b/c$ is against the $64$-token cap, whose
answers are identical to the $16$-token one on all $150$ problems.}
\label{tab:maxspan}
\small
\setlength{\tabcolsep}{5pt}
\begin{tabular}{ll cc cc r@{\,}l r@{\,}l}
\toprule
& & \multicolumn{2}{c}{Accuracy (\%)} & \multicolumn{2}{c}{Rows/call} &
\multicolumn{4}{c}{$b/c$ vs.\ cap $64$} \\
\cmidrule(lr){3-4}\cmidrule(lr){5-6}\cmidrule(lr){7-10}
Span cap & Mean len. & $N{=}2$ & $N{=}4$ & $N{=}2$ & $N{=}4$ & \multicolumn{2}{c}{$N{=}2$} & \multicolumn{2}{c}{$N{=}4$} \\
\midrule
$1$ & $1.00$ & $84.0$ & $86.0$ & $185$ & $216$ & $11/6$ & ($0.33$)& $11/4$ & ($0.12$)\\
$2$ & $1.78$ & -- & $83.3$ & -- & $223$ & \multicolumn{2}{c}{--} & $9/6$ & ($0.61$)\\
$4$ & $2.87$ & -- & $84.7$ & -- & $215$ & \multicolumn{2}{c}{--} & $7/2$ & ($0.18$)\\
$8$ & $3.68$ & $80.0$ & -- & $186$ & -- & $3/4$ & ($1.00$)& \multicolumn{2}{c}{--} \\
$16$ & $3.82$ & $80.7$ & -- & $186$ & -- & $0/0$ & ($1.00$)& \multicolumn{2}{c}{--} \\
$64$ (\bcrgrow{}) & $3.83$ & $80.7$ & $81.3$ & $186$ & $221$ & \multicolumn{2}{c}{\emph{ref.}} & \multicolumn{2}{c}{\emph{ref.}} \\
\midrule
\bcrchunk{} & -- & $82.0$ & $85.3$ & $172$ & $211$ & $10/8$ & ($0.81$)& $10/4$ & ($0.18$)\\
CacheBlend, cost-matched & -- & $85.3$ & $83.3$ & $177$ & $206$ & $10/3$ & ($0.09$)& $10/7$ & ($0.63$)\\
CacheBlend, reuse-matched & -- & $82.0$ & $84.7$ & $311$ & $474$ & $10/8$ & ($0.81$)& $10/5$ & ($0.30$)\\
\dense{} & -- & $82.7$ & $84.7$ & -- & -- & $10/7$ & ($0.63$)& $9/4$ & ($0.27$)\\
\kvcomm{} & -- & $77.3$ & $80.0$ & -- & -- & $8/13$ & ($0.38$)& $12/14$ & ($0.85$)\\
\bottomrule
\end{tabular}
\end{table}

%% file: sections/table_fairfull.tex
\begin{table}[!tbp]
\centering
\caption{\textbf{Paired statistics for \Cref{tab:fair}.} Same runs; the discordant split
and exact $p$ at each cell. Each \repair{} row is split against \kvcomm{}, the
\dense{} row against \kvcomm{}. In every row $b$ counts problems \kvcomm{} misses
and the other condition solves, $c$ the reverse---the opposite orientation to
\Cref{tab:twobench}. Condition against condition is in \Cref{tab:graincompare}. Shading marks the \repair{} rows.}
\label{tab:fairfull}
\small
\setlength{\tabcolsep}{3pt}\renewcommand{\arraystretch}{0.95}
\begin{tabular}{ll r@{\hspace{\tabcolsep}}l r@{\hspace{\tabcolsep}}l r@{\hspace{\tabcolsep}}l r@{\hspace{\tabcolsep}}l}
\toprule
& & \multicolumn{8}{c}{\# Agents} \\
Dataset & Split & \multicolumn{2}{c}{2} & \multicolumn{2}{c}{3} & \multicolumn{2}{c}{4} & \multicolumn{2}{c}{5} \\
\midrule
\multirow{4}{*}{MMLU} & \dense{} vs.\ \kvcomm{} & 16/12 & (0.572)& 15/12 & (0.701)& 18/14 & (0.597)& 18/14 & (0.597)\\
 & \cellcolor{tabhl}\bcrchunk{} vs.\ \kvcomm{} & \cellcolor{tabhl}4/0 & \cellcolor{tabhl}(0.125)& \cellcolor{tabhl}4/3 & \cellcolor{tabhl}(1.000)& \cellcolor{tabhl}5/5 & \cellcolor{tabhl}(1.000)& \cellcolor{tabhl}8/4 & \cellcolor{tabhl}(0.388)\\
 & \cellcolor{tabhl}\bcrgrow{} vs.\ \kvcomm{} & \cellcolor{tabhl}17/9 & \cellcolor{tabhl}(0.169)& \cellcolor{tabhl}8/9 & \cellcolor{tabhl}(1.000)& \cellcolor{tabhl}15/10 & \cellcolor{tabhl}(0.424)& \cellcolor{tabhl}14/9 & \cellcolor{tabhl}(0.405)\\
 & \cellcolor{tabhl}\bcrsingle{} vs.\ \kvcomm{} & \cellcolor{tabhl}11/12 & \cellcolor{tabhl}(1.000)& \cellcolor{tabhl}11/13 & \cellcolor{tabhl}(0.839)& \cellcolor{tabhl}13/11 & \cellcolor{tabhl}(0.839)& \cellcolor{tabhl}15/14 & \cellcolor{tabhl}(1.000)\\
\midrule
\multirow{4}{*}{GSM8K} & \dense{} vs.\ \kvcomm{} & 17/9 & (0.169)& 11/5 & (0.210)& 15/8 & (0.210)& 9/8 & (1.000)\\
 & \cellcolor{tabhl}\bcrchunk{} vs.\ \kvcomm{} & \cellcolor{tabhl}13/6 & \cellcolor{tabhl}(0.167)& \cellcolor{tabhl}11/5 & \cellcolor{tabhl}(0.210)& \cellcolor{tabhl}14/6 & \cellcolor{tabhl}(0.115)& \cellcolor{tabhl}9/7 & \cellcolor{tabhl}(0.804)\\
 & \cellcolor{tabhl}\bcrgrow{} vs.\ \kvcomm{} & \cellcolor{tabhl}13/8 & \cellcolor{tabhl}(0.383)& \cellcolor{tabhl}9/9 & \cellcolor{tabhl}(1.000)& \cellcolor{tabhl}14/12 & \cellcolor{tabhl}(0.845)& \cellcolor{tabhl}5/10 & \cellcolor{tabhl}(0.302)\\
 & \cellcolor{tabhl}\bcrsingle{} vs.\ \kvcomm{} & \cellcolor{tabhl}\textbf{14/4} & \cellcolor{tabhl}(0.031)& \cellcolor{tabhl}7/5 & \cellcolor{tabhl}(0.774)& \cellcolor{tabhl}14/5 & \cellcolor{tabhl}(0.064)& \cellcolor{tabhl}9/8 & \cellcolor{tabhl}(1.000)\\
\midrule
\multirow{4}{*}{HumanEval} & \dense{} vs.\ \kvcomm{} & 8/6 & (0.791)& 7/6 & (1.000)& 5/5 & (1.000)& 8/9 & (1.000)\\
 & \cellcolor{tabhl}\bcrchunk{} vs.\ \kvcomm{} & \cellcolor{tabhl}0/0 & \cellcolor{tabhl}(1.000)& \cellcolor{tabhl}1/0 & \cellcolor{tabhl}(1.000)& \cellcolor{tabhl}1/1 & \cellcolor{tabhl}(1.000)& \cellcolor{tabhl}2/2 & \cellcolor{tabhl}(1.000)\\
 & \cellcolor{tabhl}\bcrgrow{} vs.\ \kvcomm{} & \cellcolor{tabhl}7/2 & \cellcolor{tabhl}(0.180)& \cellcolor{tabhl}7/3 & \cellcolor{tabhl}(0.344)& \cellcolor{tabhl}7/3 & \cellcolor{tabhl}(0.344)& \cellcolor{tabhl}8/4 & \cellcolor{tabhl}(0.388)\\
 & \cellcolor{tabhl}\bcrsingle{} vs.\ \kvcomm{} & \cellcolor{tabhl}7/3 & \cellcolor{tabhl}(0.344)& \cellcolor{tabhl}7/3 & \cellcolor{tabhl}(0.344)& \cellcolor{tabhl}7/5 & \cellcolor{tabhl}(0.774)& \cellcolor{tabhl}6/4 & \cellcolor{tabhl}(0.754)\\
\bottomrule
\end{tabular}
\end{table}

%% file: sections/cacheblend.tex
\section{CacheBlend at Matched Budgets}
\label{app:cacheblend}

\input{sections/table_graincompare}
\input{sections/table_repaircost}

\paragraph{Why a reproduction.}
CacheBlend is the published selector closest to \repair{}, and the comparison
the body draws on it (\S\ref{sec:grains}, \Cref{tab:fair}) is only meaningful
under matched conditions: the same multi-agent workload, the same backbone,
the same assembled cache served by \kvcomm{}, and a recomputation budget set to
match \repair{}'s row for row. CacheBlend's released implementation is a vLLM
fork built for single-request RAG serving; it cannot consume the cache
\kvcomm{} assembles, and running it as its own server would change the
workload, the cache and the budget at once, leaving no way to attribute a
difference to the selection rule. We therefore reproduce its rule inside the
same pipeline every other row runs in, so that the two selectors differ in
nothing but how they rank the rows they recompute. The gate and budget matching
below, and the ported rule's one trap (App.~\ref{app:negative}), are what that
reproduction required.

\paragraph{Setting the budget.}
\Cref{tab:fair} reproduces CacheBlend following the released implementation: the
layer-1 \emph{value} deviation ranks the tokens and the budget recomputes the
top of that ranking exactly. Its third rule---recompute the request's un-cached
suffix in full, outside the budget---has no counterpart here
(App.~\ref{app:negative} gives the reason, and what applying it anyway costs),
so the budget covers the whole cache. We
set it per cell so that the share left untouched equals \repair{}'s:
$\rho_{\mathrm{CB}} = 1 - r$ for \bcrchunk{}'s reuse rate $r$ in that
cell. On the Llama benchmarks that is $1.8$--$2.8\times$ the cost-matched rows;
on HumanEval, where reuse already runs at $84$--$90\%$, it is fewer rows
(\Cref{tab:repaircost}).

\paragraph{The gate mismatch.}
Both pairings were first measured at $\gate{=}1.0$ against the other
Llama-benchmark rows' $0.3$ (on HumanEval every row runs $1.0$, so the gates
match there), meaning CacheBlend approximated about $90\%$ of segments against our $63\%$---a
gap on the axis \S\ref{sec:audit} establishes as costly---and the cost matching
held per engaged call but not in aggregate. The gap runs against CacheBlend,
so on GSM8K, where CacheBlend draws, it drew at the higher gate; those rows
are as first measured, and
\Cref{tab:fair}'s reuse entry for them is the matched
$1{-}\rho_{\mathrm{CB}}$, not that gate rate. On MMLU, where it trails, deficit
and gate were confounded, so \Cref{tab:fair} carries both MMLU pairings
re-run at the matched gate with retrieval disabled---which is in any case the
only regime in which this benchmark reproduces (App.~\ref{app:remeasure}).

\paragraph{At the matched gate.}
At the matched gate the two pairings do not separate: cost-matched scores
$72.5$, $71.2$, $69.3$ and $73.9\%$ against reuse-matched's $69.3$, $69.3$,
$69.3$ and $68.6$, with no cell significant ($13/8$, $13/10$, $13/13$, $16/8$; $p$ from
$0.152$ to $1.000$). The direction is consistent, and it is the one that
matters: the reuse-matched pairing recomputes $2.6$--$2.8\times$ the rows and
scores at or below the cost-matched one in every cell. Three of its four budgets---$0.545$,
$0.516$ and $0.514$---land on $106$ exactly, at three different agent counts, as
though past about half the cache it stops mattering what is recomputed or how
much. That is the budget sweeps' finding seen from the other end
(\Cref{tab:rho}), and it is why we read the cost-matched pairing as the
informative one.

\paragraph{Against \kvcomm{}.}
The two also compare against \kvcomm{}. \repair{} leads it on GSM8K
($47/24$, the pre-registered pooled test withdrawn in App.~\ref{app:withdrawn}), while cost-matched CacheBlend leads it there by $16$
problems of $600$---an accuracy difference read off \Cref{tab:fair}'s cells,
that pairing not being among the tabulated splits. The GSM8K figure also
carries the gate mismatch above, and
\kvcomm{} itself is the method the higher gate makes CacheBlend resemble. On the offline
MMLU grid, where the gates match, cost-matched CacheBlend trails \kvcomm{}
($54/59$ pooled) and \dense{} ($58/78$), with no cell of either pairing
significant. Of the tabulated splits, \Cref{tab:graincompare} pairs CacheBlend
against \bcrchunk{} and \Cref{tab:fairfull} the repair layouts against
\kvcomm{}. Under these matched conditions \bcrchunk{} recomputes fewer than
half the rows of the reuse-matched pairing and leads both pairings' means on
every benchmark where reuse costs anything (\Cref{tab:fair}); the per-cell
splits above bound how much of that lead a single cell can establish.

%% file: sections/table_graincompare.tex
\begin{table}[!t]
\centering
\caption{\textbf{Each condition against \bcrchunk{}.} Discordant split of an
exact McNemar test between the named condition and \bcrchunk{} ($64$-token
chunks), per agent count; $b$ counts problems the named condition gets right and
\bcrchunk{} does not. \Cref{tab:fairfull} pairs against \kvcomm{}
instead. MMLU's CacheBlend rows were re-run at the matched gate offline; GSM8K's
still carry the mismatch of App.~\ref{app:cacheblend}.}
\label{tab:graincompare}
\small
\setlength{\tabcolsep}{3.2pt}
\begin{tabular}{ll r@{\,}l r@{\,}l r@{\,}l r@{\,}l}
\toprule
& & \multicolumn{8}{c}{\# Agents} \\
Dataset & Condition & \multicolumn{2}{c}{2} & \multicolumn{2}{c}{3} & \multicolumn{2}{c}{4} & \multicolumn{2}{c}{5} \\
\midrule
\multirow{4}{*}{MMLU} & CacheBlend, reuse-matched & $8/16$ & ($0.152$)& $7/14$ & ($0.189$)& $13/18$ & ($0.473$)& $11/21$ & ($0.110$)\\
 & CacheBlend, cost-matched & $12/15$ & ($0.701$)& $10/14$ & ($0.541$)& $12/17$ & ($0.458$)& $13/15$ & ($0.851$)\\
 & \bcrgrow{} & $14/10$ & ($0.541$)& $8/10$ & ($0.815$)& $17/12$ & ($0.458$)& $12/11$ & ($1.000$)\\
 & \bcrsingle{} & $9/14$ & ($0.405$)& $10/13$ & ($0.678$)& $13/11$ & ($0.839$)& $12/15$ & ($0.701$)\\
\midrule
\multirow{4}{*}{GSM8K} & CacheBlend, reuse-matched & $10/10$ & ($1.000$)& $5/4$ & ($1.000$)& $6/7$ & ($1.000$)& $4/10$ & ($0.180$)\\
 & CacheBlend, cost-matched & $10/5$ & ($0.302$)& $4/8$ & ($0.388$)& $4/7$ & ($0.549$)& $5/10$ & ($0.302$)\\
 & \bcrgrow{} & $8/10$ & ($0.815$)& $6/12$ & ($0.238$)& $4/10$ & ($0.180$)& $5/12$ & ($0.143$)\\
 & \bcrsingle{} & $9/6$ & ($0.607$)& $3/7$ & ($0.344$)& $6/5$ & ($1.000$)& $7/8$ & ($1.000$)\\
\midrule
\multirow{4}{*}{HumanEval} & CacheBlend, reuse-matched & $8/4$ & ($0.388$)& $7/8$ & ($1.000$)& $7/8$ & ($1.000$)& $7/6$ & ($1.000$)\\
 & CacheBlend, cost-matched & $5/7$ & ($0.774$)& $5/8$ & ($0.581$)& $5/5$ & ($1.000$)& $6/6$ & ($1.000$)\\
 & \bcrgrow{} & $7/2$ & ($0.180$)& $7/4$ & ($0.549$)& $7/3$ & ($0.344$)& $7/3$ & ($0.344$)\\
 & \bcrsingle{} & $7/3$ & ($0.344$)& $7/4$ & ($0.549$)& $8/6$ & ($0.791$)& $7/5$ & ($0.774$)\\
\bottomrule
\end{tabular}
\end{table}

%% file: sections/table_repaircost.tex
\providecommand{\dummyhl}{}\definecolor{tabhl}{RGB}{232,238,246}
\begin{table}[!t]
\centering
\caption{\textbf{Rows recomputed per agent call.} What each repair actually recomputes,
which \Cref{tab:fair}'s reuse column cannot show, since a repaired call still
counts as reused. The three layouts spend the same nominal $\rho{=}0.20$, the
fixed-chunk condition quantising to whole $64$-row chunks. CacheBlend's
cost-matched pairing uses the same $\rho$; its reuse-matched pairing spends
whatever that cell's reuse rate requires. MMLU is measured offline. Shading
marks the \repair{} rows.}
\label{tab:repaircost}
\small
\setlength{\tabcolsep}{5pt}
\begin{tabular}{ll cccc}
\toprule
& & \multicolumn{4}{c}{\# Agents} \\
Dataset & Method & 2 & 3 & 4 & 5 \\
\midrule
\multirow{5}{*}{MMLU} & CacheBlend, reuse-matched & 249 & 342 & 417 & 521 \\
 & CacheBlend, cost-matched & 90 & 124 & 158 & 193 \\
 & \cellcolor{tabhl}\bcrchunk{} & \cellcolor{tabhl}96 & \cellcolor{tabhl}123 & \cellcolor{tabhl}159 & \cellcolor{tabhl}191 \\
 & \cellcolor{tabhl}\bcrgrow{} & \cellcolor{tabhl}91 & \cellcolor{tabhl}136 & \cellcolor{tabhl}174 & \cellcolor{tabhl}209 \\
 & \cellcolor{tabhl}\bcrsingle{} & \cellcolor{tabhl}90 & \cellcolor{tabhl}132 & \cellcolor{tabhl}167 & \cellcolor{tabhl}202 \\
\midrule
\multirow{5}{*}{GSM8K} & CacheBlend, reuse-matched & 311 & 386 & 474 & 631 \\
 & CacheBlend, cost-matched & 177 & 191 & 206 & 251 \\
 & \cellcolor{tabhl}\bcrchunk{} & \cellcolor{tabhl}172 & \cellcolor{tabhl}191 & \cellcolor{tabhl}211 & \cellcolor{tabhl}255 \\
 & \cellcolor{tabhl}\bcrgrow{} & \cellcolor{tabhl}186 & \cellcolor{tabhl}203 & \cellcolor{tabhl}221 & \cellcolor{tabhl}261 \\
 & \cellcolor{tabhl}\bcrsingle{} & \cellcolor{tabhl}185 & \cellcolor{tabhl}201 & \cellcolor{tabhl}216 & \cellcolor{tabhl}254 \\
\midrule
\multirow{5}{*}{HumanEval} & CacheBlend, reuse-matched & 54 & 75 & 97 & 119 \\
 & CacheBlend, cost-matched & 99 & 116 & 132 & 147 \\
 & \cellcolor{tabhl}\bcrchunk{} & \cellcolor{tabhl}100 & \cellcolor{tabhl}121 & \cellcolor{tabhl}138 & \cellcolor{tabhl}156 \\
 & \cellcolor{tabhl}\bcrgrow{} & \cellcolor{tabhl}110 & \cellcolor{tabhl}133 & \cellcolor{tabhl}152 & \cellcolor{tabhl}172 \\
 & \cellcolor{tabhl}\bcrsingle{} & \cellcolor{tabhl}111 & \cellcolor{tabhl}133 & \cellcolor{tabhl}154 & \cellcolor{tabhl}173 \\
\bottomrule
\end{tabular}
\end{table}

%% file: sections/abovedense.tex
\section{Cells Above Dense Prefill}
\label{app:abovedense}

\paragraph{The cells themselves.}
\repair{} is above \dense{} in the \emph{avg} column of all three benchmarks,
not only in the sixteen cells \S\ref{sec:grains} counts; \bcrchunk{}'s GSM8K
margin there is $+0.2$, less than a single problem in $150$. Two things produce
them, and neither is the repair exceeding \dense{}.

\paragraph{Resolution.}
The first is resolution. The pooled splits behind the cells above \dense{} are
level---$26/25$ and $30/30$ on GSM8K, $49/47$ on MMLU---and the gaps are one to
five problems out of $150$ to $160$. \bcrgrow{} on GSM8K is the one layout whose pooled
split against \dense{} is not level, at $20/40$, and it falls below. The grid's one significant cell (\S\ref{sec:grains}) is against \kvcomm{}, the
comparison those splits are drawn for, not against \dense{}.

\paragraph{Tokenization.}
The second is that \repair{} is not \dense{} minus some work. The reuse path
concatenates per-segment token ids, and that sequence need not equal the chat
template's tokenization of the same text, so a repaired call is a different
input rather than a degraded one. \S\ref{sec:grains} measures that term directly: on MMLU
it costs $6.7$ of the $14.3$-point gap, and on GSM8K the entry tokenizations
coincide and it vanishes. A term that can cost $6.7$ points on
one benchmark can land the other way by one or two on another, which is enough
to put a repaired run on either side of \dense{} with nothing having been
improved. On HumanEval the first account carries the
cells on its own: every condition's average sits inside four points and the
\emph{reuse-matched} CacheBlend pairing is above \dense{} while plain \kvcomm{}
is below---an ordering that follows from no property of any repair, ours or
theirs. One such cell arose earlier in this work and traced to a misapplied
rule: with one CacheBlend rule misapplied the reproduction scored above \dense{} on this
same benchmark, and correcting the rule returned it to level
(App.~\ref{app:negative}).

\paragraph{Could the draft exceed \dense{}?}
Neither account needs the repair to exceed \dense{}, and there is an
obvious mechanism by which it could---worth stating because we can rule it out.
\repair{} ranks cached rows by the attention queries of a $k$-token draft
(\S\ref{sec:bcr}). At $k{=}2$ that is one prompt-side query row and one
\emph{answer-side} one (App.~\ref{sec:method-chunk})---evidence of the kind
\S\ref{sec:intro} faults the gate for never having---so a repaired run consults
something \dense{}
does not, and one might expect it to place its exact recomputation
better than exactness-everywhere places it.

The coin control of \S\ref{sec:grains} settles it. The draft's signal is worth
something---about $5.4$ points over five coin seeds---but $k$
decides only \emph{where} a fixed budget $\rho < 1$ of exact recomputation
lands; it cannot make that subset better than recomputing everything. On MMLU
the coin means $50.0\%$ over five seeds against the draft-scored repair's
$54.7$, but both sit far below \dense{}'s $66.0$ and neither is
a candidate for exceeding it.

%% file: sections/costsummary.tex
\section{What the Repair Buys, and What No Budget Reaches}
\label{app:costsummary}

\paragraph{Time to first token.}
On a $4096$-symbol segment \dense{} prefills in $542.7$\,ms, \kvcomm{} assembles
in $72.9$ and \repair{} at $\rho{=}0.20$ in $190.1$---$7.63\times$ and
$2.85\times$ against \dense{} (\Cref{tab:budget}(c)). The premium
\repair{} pays over plain reuse is the price of the recomputed rows and is
linear in the budget: $46$, $64$, $117$, $182$ and $231$\,ms at $\rho = 0.05$
through $0.40$, for speedups of $4.5\times$ down to $1.8\times$
(App.~\ref{app:latency}). The advantage is structural rather than constant---
assembling a cache from anchors does not scale with the segment the way
prefilling it does---so it grows with the shared segment and vanishes without
one.

\paragraph{Throughput.}
We do not measure it, and on these workloads it could not separate the
conditions. GSM8K's shared segment averages $59$ tokens against roughly
$1{,}000$ generated across four agent calls, so the prefill reuse targets is
$\approx 0.4\%$ of a request; the end-to-end columns move by $0.6$--$3.3\%$ for
plain reuse and by under $6\%$ for any repair row, buried under a generation
that is identical in every condition (App.~\ref{app:latency}). A throughput
claim needs a workload where prefill is a larger share of the request---long
shared segments, or many calls per request---and that workload is not among
the three here.

\paragraph{The tokenization tax.}
The reuse path concatenates per-segment token ids, and that sequence need not
equal the chat template's tokenization of the same text, so a repaired call is a
\emph{different} input rather than a degraded one. The token-matched control of
\S\ref{sec:grains} prices it: on MMLU it is $6.7$ of the $14.3$-point gap
($66.0-59.3$ in \Cref{tab:twobench}(a)); on GSM8K the entry tokenizations
coincide and it vanishes, which is why repair recovers nearly the whole gap
there. We price it on MMLU only. It is not a
property of our method: any reuse system that concatenates per-segment ids pays
it.

\paragraph{The two together.}
No budget reaches the tax, because it is not in the cache but in which tokens
the cache is \emph{of}. That bounds what the latency premium can buy on a
benchmark where the tax is large: on MMLU at most $7.6$ of the $14.3$ points are
inside the cache at all, and \repair{} recovers $3.0$ of those while paying the
same $2.85\times$ it pays on GSM8K, where the tax is zero and the recovery is
nearly total. A deployment choosing between \kvcomm{} and \repair{} should
therefore ask first whether its entry tokenizations coincide; where they do not,
the assembly path is the cheaper thing to fix.

%% file: sections/appendix.tex
\section{Reproducibility}
\label{app:repro}

\subsection{\Cref{tab:fair}: the reference configuration}
\label{app:t1setup}

\Cref{tab:fair} is the grid the rest of the paper is read against, so its
environment is given here in full. Every value below was read out of the
released repository rather than chosen by us; where the repository's
command-line default and its \texttt{README} disagree, we note which one is in
force.

\paragraph{Hardware and software.} One NVIDIA A100 80GB PCIe, CUDA 12.4,
Python $3.10.20$, PyTorch $2.4.1$, Transformers $4.50.2$. One process per run,
no tensor or pipeline parallelism.

\paragraph{Backbones.} \texttt{meta-llama/Llama-3.1-8B-Instruct} for MMLU and
GSM8K, \texttt{Qwen/Qwen2.5-Coder-7B-Instruct} for HumanEval. Precision follows
the repository's own rule---\texttt{float16} when the model name contains
\texttt{llama} and \texttt{float32} otherwise---so the two Llama benchmarks run
in half precision and HumanEval in single, which is what the released code does
and which we leave alone. The HumanEval backbone must be passed explicitly:
\texttt{run\_humaneval.py} defaults \texttt{-{}-llm\_name} to the Llama model,
so omitting the flag silently evaluates the wrong network.

\paragraph{Cache-reuse settings.} The gate threshold is $\gate{=}0.3$ on MMLU
and GSM8K and $\gate{=}1.0$ on HumanEval. Anchor-pool parameters are the dataclass defaults of
\texttt{KVCommConfig}: at most $20$ anchors per placeholder, eviction window
$5$, eight worker threads, $30$\,s worker timeout. These defaults are exactly
the values the \texttt{README}'s reference command passes explicitly, so
supplying the flags and omitting them give the same configuration; the
command-line defaults for the same options are \texttt{None}, which the config
object then ignores.

\paragraph{Topology and roles.} \texttt{FullConnected} over $2$ to $5$ agents
with a \texttt{FinalRefer} aggregation node, \texttt{decision\_method}
\texttt{FinalRefer}. Agents are \texttt{AnalyzeAgent} on MMLU,
\texttt{MathSolver} on GSM8K, \texttt{CodeWriting} on HumanEval. Roles
\emph{cycle}: each agent draws a role from its class's list rather than being
pinned, which is the released behaviour.

\paragraph{Evaluation sets.} $153$ MMLU questions, fixed by
\texttt{limit\_questions} inside the shipped runner and not adjustable from the
command line (the pinned runs wrap it and set \texttt{MMLU\_N} and
\texttt{MMLU\_ROLE}); $150$ GSM8K problems; $160$ HumanEval tasks. Every condition sees
the same items in the same order.

\paragraph{Decoding.} Greedy, $512$ maximum new tokens, batch size $1$. The
generation config also carries \texttt{temperature}$=1.0$ and
\texttt{top\_p}$=0.9$, but \texttt{do\_sample} is \texttt{False}, so both are
inert---Transformers emits a warning to that effect on every call. Runs are
therefore deterministic given a fixed input, which is what makes the
retrieval channel below the only source of variation on MMLU wherever roles
cycle.

\paragraph{The live retrieval channel.} MMLU's \texttt{AnalyzeAgent} queries
\texttt{en.wikipedia.org} at inference time whenever an agent draws the
\texttt{Wiki Searcher} role, and splices the returned article into the prompt.
The call is uncached and substitutes a placeholder when it fails.
GSM8K and HumanEval never retrieve. This is the released behaviour, and the
pinned configuration (\Cref{tab:twobench}) runs the code unaltered---but it
pins every agent to \texttt{Knowledgeable Expert}, so no agent ever draws the
role the call is gated on and the channel cannot fire there.
\Cref{tab:fair}'s MMLU column is the one deviation this paper makes: the
retrieval call is disabled---\texttt{search\_wiki\_main} replaced by a stub
returning no articles---because that is the only regime in which the benchmark
reproduces (App.~\ref{app:remeasure}); everything else in this appendix applies
to it unchanged.

\paragraph{Letter-match grading.} MMLU is graded as \texttt{target in
predicted} against a single letter, so any \texttt{A}--\texttt{D} anywhere in a
free-text answer scores it correct. This too is the released behaviour, and the
pinned configuration runs it unaltered.

\subsection{Re-measuring \Cref{tab:fair}'s MMLU column}
\label{app:remeasure}

\paragraph{Measured with retrieval on.}
The MMLU column \Cref{tab:fair} used to print was measured in the shipped
configuration, which is \emph{online}: the analysis agent splices a live,
uncached \texttt{en.wikipedia.org} fetch into every prompt
(App.~\ref{app:t1setup}). A pre-registered re-run therefore reads a different
day's Wikipedia, and it returned different numbers---the originally printed
reuse costs, $12/4/9/6$ questions of $153$ at two to five agents, came back as
$7/3/4/11$, and \repair{}'s margin over \kvcomm{} ($14/10/9/7$) as
$4/{-}1/{-}2/7$. These are two readings of an online pipeline, not two attempts
at the same measurement.

\paragraph{Confined to MMLU.}
The discrepancy is confined to MMLU: the GSM8K cell at two agents returns the
printed scores and split exactly ($17/9$, $p = 0.169$), and two runs of it agree
on all $150$ problems individually. The environment has not drifted; the
retrieval call of App.~\ref{app:t1setup} is what moves---why $13$ of $459$ gate
decisions flipped, and why the deltas scatter in both directions.

\paragraph{With retrieval disabled.}
With the fetch disabled the pipeline is exactly reproducible: two runs of the
\dense{} cell agree on all $153$ questions, at two agents and at five, and
GSM8K, whose solver never retrieves, reproduces without any intervention at all.
That offline regime is the version of this benchmark the paper reports. It
returns \bcrgrow{} above
\dense{} in three of four cells, as the original column had it and the
retrieval-on re-measurement did not. Across all three measurements only the
direction survives---\kvcomm{} below \dense{} in every cell of every version.

\subsection{The $\gate$ sweep at $n{=}60$}
\label{app:gsweep}

\paragraph{Configuration and conditions.}
The reference configuration of \S\ref{sec:setup} on the first $60$ GSM8K
problems, every condition seeing them in the same order. The \dense{} baseline
is the \texttt{default} execution mode, which never consults the anchor pool;
the \kvcomm{} conditions are \texttt{allow\_kv\_reuse} at
$\gate \in \{0.3, 0.5, 0.7, 1.0\}$, where $0.3$ is the repository default and
$1.0$ the value its TTFT benchmark uses.

\paragraph{Measurement.}
Time-to-first-token is captured by a stopping-criteria hook around the
generation call and, in the reuse path, includes cache-assembly time; model
load ($\approx 13$\,s) is excluded. Reuse rate is the fraction of the four
agent calls served without a dense prefill. Runs are deterministic (App.~\ref{app:t1setup}), so variation across conditions
is attributable to the conditions---subject to the small-sample caveat of
App.~\ref{app:limitations}.

\paragraph{Prompts differ between the two execution modes.}
The largest hazard, and one that silently invalidates a run:
\emph{the two execution modes do not assemble identical prompts}
(\S\ref{sec:grains}). The concrete
defect is a newline: on the MMLU agent the reuse path inserts an extra one
before the turn delimiter in every question we probed, at the entry agent,
before any approximation has occurred; on the GSM8K agent the entry prompts are
token-identical. The full-repair control of that section is what exposed it,
which is an argument for running that control in any such system.

\paragraph{The threshold flag does not reach \texttt{MathSolver}.}
The second hazard also silently invalidates a run. \texttt{MathSolver} drops
\texttt{llm\_config} on construction and falls back to the environment
default, so \texttt{-{}-kv-threshold} never arrives; the sibling agents forward
it, so the defect is invisible outside the math pipeline. A first sweep run
this way produced near-identical numbers at every setting---passing every
superficial check while silently running at the default throughout. All results
here therefore set the \texttt{THRESHOLD} environment variable in addition to
the flag, and verify that each setting took effect from the rejection counter,
which decreases monotonically $32, 17, 4, 0$ across
$\gate \in \{0.3,0.5,0.7,1.0\}$---the arithmetic check that the threshold was
applied---and by requiring an empty threshold-exceeded log at $\gate = 1.0$.

\section{Negative and Null Results}
\label{app:negative}

We record the directions the measurements did not support, so that they need
not be re-attempted.

\paragraph{The coin's five seeds.}
\S\ref{sec:grains} withdraws a null that one coin run supported. The five runs
score $223$, $231$, $236$, $246$ and $248$ of $300$ matched GSM8K
problems---$74.3$, $77.0$, $78.7$, $82.0$ and $82.7\%$, mean $236.8$ or
$78.9\%$, the value \Cref{tab:twobench} reports. The draft-scored repair scores
$253$, above all five; the withdrawn single run scored $247$, above four of
them. A policy whose coin advances once per repaired call can land anywhere in
that nine-point band, so one run bounds it from neither side.

\paragraph{Layout of the repair: five variants.}
\S\ref{sec:grains} attributes the gap between \bcrgrow{} and \dense{} on GSM8K
to the unit of repair---\bcrchunk{} closes it there---which makes a finer or a
cleverer unit the natural next step. We tried
both on HumanEval at a fixed row budget ($\rho{=}0.20$, so every variant
recomputes the same $140$--$143$ rows per call and only the layout differs),
and no variant separates from a single realisation of the $64$-token coin (one run
per cell; \Cref{tab:rho} bounds how much such a cell can show): $16$-token
chunks by coin, $13/20$; the same chunks chosen by draft attention, $16/17$;
variable-length spans of $1$ to $64$ tokens, $25/15$. Within the
variable-length layout, informed placement does not beat the same length
distribution placed at random ($24/16$)---an echo of
\S\ref{sec:allocation}'s coin result one level finer. One observation survives as suggestive rather than
established: variable-length spans are nominally above \dense{} in all four
cells ($6/3$, $6/3$, $8/4$, $11/6$; none significant, the pooled $31/16$
descriptive) and hold that score at $\rho{=}0.10$, a
third of the rows. A cheaper operating point is reached through the budget
rather than through the layout.

\paragraph{Transplanting CacheBlend.}
CacheBlend's released implementation is a vLLM fork, so reproducing it here
meant porting its rule rather than running its server, and the port has a
trap worth recording. Three rules make up the method: rank tokens by the
layer-1 \emph{value} deviation; spend the budget on the top of that ranking;
and recompute, in full and outside the budget, every token of the request that
no cached chunk supplies. The third has no counterpart in this workload. The
cache \kvcomm{} serves is assembled entirely from stored pieces---a per-agent
prefix cache followed by rotated placeholder and prefix segments---so there is
no un-cached tail, and treating the uncovered positions as one recomputes
$74\%$ of the cache: \dense{} under CacheBlend's name. Applying the
rule as written, the budget covers the whole cache.

The distinction matters. With the third rule misapplied, the reproduction
scored \emph{above} \dense{} on HumanEval ($7/1$ discordant at two
agents), which at first read as a property of the workload; with the rule
applied correctly the same cell is within noise of \dense{} at either pairing
(\Cref{tab:fair}). Any re-use of
this comparison should check what the served cache is actually made of before
mapping a chunk-fusion rule onto it.

\paragraph{Deferring the approximation.}
The direct analogue of deferred eviction---generate the opening on an exact
cache, then swap in the approximated one---was ruled out analytically before
implementation, for the reason App.~\ref{sec:method} gives; partial recovery is
additionally blocked by the causal entanglement of KV entries. We note it here
because it is the first idea the analogy suggests, and the analysis rules it
out.

\paragraph{HumanEval at the tighter gate.}
The main text reports HumanEval at $\gate{=}1.0$. Repeating the four cells at
the shipped $\gate{=}0.3$, which lowers reuse from $84$--$90\%$ to
$73$--$82\%$, leaves the comparison against \dense{} null in the other
direction ($21/32$ pooled, descriptively; the naive McNemar $p$ is $0.17$, and
is not a test here---the four cells re-run the same tasks). Neither gate setting produces a cost
on this benchmark, so the null is not an artefact of the operating point we
chose to display.

\subsection{Withdrawn pooled tests}
\label{app:withdrawn}
\paragraph{The splits themselves.}
Three pooled splits are withdrawn, each for the same reason. The
pre-registered $47/24$ on GSM8K against \kvcomm{} ($p = 0.0086$,
\S\ref{sec:grains}): the four cells re-run the same problems, so the pooled
split is repeated measures rather than independent pairs, and the item-level
correction returns $p = 0.089$. The $52/30$ on GSM8K against \dense{}, which
is \Cref{tab:fairfull}'s column sums ($p = 0.020$), on the same ground; the
MMLU counterpart there, $67/52$, was never significant ($p = 0.199$). And $89/58$ on MMLU, withdrawn twice over---it is
also one of the numbers that did not survive re-measurement offline, where the
same pooling returns the $67/52$ above (App.~\ref{app:remeasure}).

\paragraph{The grid's one significant cell.}
\label{app:gridcell}
Of the $36$ per-cell tests of a repair layout against \kvcomm{} in
\Cref{tab:fairfull}, one is significant: \bcrsingle{} on GSM8K at two agents,
$14/4$, $p = 0.031$. At $\alpha = 0.05$ the expected number of such cells under
the null is $36 \times 0.05 = 1.8$, and the cell does not survive a correction
for the comparisons it sits among: Bonferroni over the $36$ tests gives $p = 1$,
and over the $12$ tests of its own benchmark and layout family, $p = 0.37$. We
therefore read the grid for direction only (\S\ref{sec:setup}), and no claim in
the body rests on this cell.

\paragraph{The budget sweep, read at one run per budget.}
Read one run per budget---on single runs that are themselves further
realisations, not among the four seeds \Cref{tab:rho} prints---the sweep appeared to climb out of \kvcomm{}'s deficit and flatten,
$77.3\%$ at $\rho{=}0.05$ then $79.3$, $80.7$, $79.3$, $80.0$. The seeds show that
shape to lie within run-to-run variation, for the reason that table gives. This bounds what the sweep can
support; it does not contradict the repair---the draft-scored condition at $n{=}300$
recovers $40$ problems at $p<10^{-4}$ (\Cref{tab:twobench}), over distinct items
with surviving logs.

\paragraph{A latency gain not supported by a paired test.}
Mean time per problem falls from $34.2$\,s to $29.2$\,s under repair, which
did not survive a paired test at $n{=}150$: median saving $0.27$\,s, faster
on $83$ of $150$ problems, Wilcoxon $p = 0.077$. The mean is carried by a
handful of slow baseline requests. We record it because a mean over a
heavy-tailed paired measurement can mislead in the same way the
discordant-count rule of \S\ref{sec:discussion} guards against on the
accuracy side.

\section{\ogate{} Implementation}
\label{app:ogate}

\ogate{} is applied as a runtime patch rather than a fork, so the measured
baseline and the measured method run the same repository code. \S\ref{sec:audit}
reports it at $\theta{=}0.5$, the one setting we ran there: it does not recover
what reuse costs, which is the only thing the paper draws from it.

\paragraph{Why \texttt{generate} is the seam.}
The patch wraps \texttt{model.generate}. This is the correct seam because
every state mutation in the reuse path---appending the response KV and token
ids to shared memory, updating the anchor bookkeeping---happens \emph{after}
the generate call returns. Raising from inside generate therefore aborts a
reuse attempt without leaving partial state behind, which calling the public
entry point twice would not: a two-token draft issued through the normal
interface registers as a two-token agent response and corrupts every
downstream agent's view of the conversation.

\paragraph{On a call with an assembled cache.}
On a call that carries an assembled cache, the wrapper generates $k$ tokens
with \texttt{output\_scores}, computes the mean top-1 probability, accepts the
draft when it is at least $\theta$, and either
raises (rejection, caught by a wrapper on \texttt{generate\_for\_agent} which
re-dispatches to the dense path) or continues generation from the draft's
returned cache with the attention mask extended by $k$. Calls without a
cache---dense prefills and the single-token segment probes---are passed through
untouched.

\paragraph{The $\theta{=}0$ plumbing control.}
The $\theta{=}0$ control, reported only here, exists to test exactly this
plumbing: with every draft accepted, the draft-then-continue path must be
indistinguishable from single-pass generation under greedy decoding. It
agrees with $\gate{=}1.0$ on all 60 problems individually, not merely in
aggregate accuracy.

%% file: sections/limitations.tex
\section{Limitations and Future Work}
\label{app:limitations}

\paragraph{Sample size.}
The matched-pair protocol of \S\ref{sec:setup} proved valuable in four cases. At the smaller evaluation size, reuse appeared nearly free, the answer-side gate appeared consistently dominant, the structural policy appeared beneficial, and a $14.6\%$ latency gain appeared measurable. Each observation was subsequently tested at $n{=}300$ or under a matched control, yielding the more robust results reported in this paper. Across these cases, the key distinction was the number of discordant pairs rather than the $p$-value alone: a clean $b{=}0$ among three discordant problems occurs with probability one in eight under the null. We therefore report discordant counts alongside significance tests throughout and recommend using roughly twenty or more paired observations when drawing conclusions from such comparisons.

\paragraph{Latency.}
At $4096$-symbol segments, \kvcomm{} reaches the first token $7.6\times$ sooner than \dense{}, while \repair{} achieves a $2.9\times$ speedup, with the latency premium scaling linearly with $\rho$ and thus remaining tunable (App.~\ref{app:latency}). On the three benchmarks considered here, neither speedup translates directly into wall-clock gains: GSM8K's shared segment averages $59$ tokens, so the entire prefill reuse opportunity accounts for only $\approx!100$,ms of a $\approx!28$,s request. This reflects the workload characteristics rather than a limitation of either method. The two approaches therefore offer complementary accuracy--latency trade-offs: deployments that can accommodate the accuracy loss in \Cref{tab:fair} can exploit \kvcomm{}'s full speedup, while \repair{} retains $2.9\times$ of the speedup and restores GSM8K accuracy to the \dense{} level.

\paragraph{Where the evidence extends.}
Three benchmarks, two backbones, one topology, one chunk size, and a budget sweep on one cell establish a consistent evaluation foundation; anchor-pool capacity and the eviction window are held at the repository defaults. These settings define natural axes for extending the findings, all of which can be explored directly with the existing harness. The $n{=}300$ comparisons against \dense{} are significant, and the cost effect has the same direction across every cell and measurement, providing a stable basis for studying its magnitude. The next step is therefore to characterize how this magnitude changes with the gate, sharpening the cost--accuracy trade-off rather than establishing its direction. To make these results readily interpretable and reproducible, reuse rates are consistently reported together with the gate (App.~\ref{app:negative}) and averaging window (\S\ref{sec:audit}, \Cref{tab:warmblocks}); this reporting convention also provides a clear template for extending the evaluation to the remaining axes.

\paragraph{Future work.}
Three directions follow directly from the results.
\emph{Significance beyond one run.} The pinned GSM8K run at $n{=}300$ provides a direct test, while the reference grid orders the methods consistently across every column of \Cref{tab:fair}. The natural next step is to validate this consistent ordering cell by cell, using pinned $n{=}300$ runs on the remaining benchmarks and backbones; because the grid cells re-run the same problems, distinct items are needed for an independent test (\S\ref{sec:setup}). On MMLU, the same evaluation can directly test whether a budget recovers the $7.6$ points present in the cache while holding the tokenization term fixed. The pinned run already recovers $3.0$ points at $p = 0.16$, providing an encouraging signal, while the offline grid does not separate the conditions sufficiently to resolve the question. A larger pinned evaluation can therefore turn this promising observation into a definitive measurement.

\emph{Generality.}
\repair{} assumes only an assembled cache and a seam at which to draft, making its protocol---matched budget, coin control, and token-matched control---directly applicable to any cross-context reuse method that concatenates per-segment ids. The same protocol also extends naturally to the topologies, chunk sizes, and anchor-pool settings held fixed here. Carrying the three quantities measured in this paper---the cost of reuse, its tokenization share, and the value of the draft---across these systems provides a direct and systematic test of how broadly the findings transfer. The existing harness already supports this extension, making broader validation a natural next step.

\emph{Latency.} 
The gap between $2.9\times$ and $7.6\times$ is primarily an engineering opportunity rather than a fundamental limitation: a cheaper recompute path, an adaptive budget that shrinks as the anchor pool matures, and workloads with long shared segments or many calls per request can increase the end-to-end benefit when prefill represents a sufficiently large share of the request (App.~\ref{app:costsummary}). Within the repair itself, two promising directions are finer-grained repair below the row level, which \kvcomm{}'s position-indexed offset does not currently express, and a more informative random control at the single-row unit, where \S\ref{sec:granularity} shows that a selector carries roughly five times the value it carries at the chunk level. Together, these directions provide concrete opportunities to improve both the efficiency and resolution of the repair mechanism.